\documentclass[twocolumn]{aastex631}
\usepackage[T1]{fontenc}
\usepackage{xspace}
\usepackage{multirow}
\usepackage{amsmath}
\usepackage{lipsum}
\usepackage{soul}
\usepackage[normalem]{ulem}
\usepackage{float}
\usepackage{comment}

\usepackage{graphicx} 

\newcommand{\nustar}{\textit{NuSTAR}\xspace}
\newcommand{\xmm}{\textit{XMM-Newton}\xspace}
\newcommand{\nicer}{\textit{NICER}\xspace}

\newcommand{\src}{PSR~J0437--4715\xspace}
\newcommand{\psrdz}{PSR~J0030+0451\xspace}
\newcommand{\psrtt}{PSR~J1231--1411\xspace}
\newcommand{\psrzs}{PSR~J0740+6620\xspace}

\shorttitle{Information Content of X-ray Pulse Profiles}
\shortauthors{G\"uver et al.}
\graphicspath{{./}{}}

\begin{document}

\title{Limits to Extracting Neutron-Star Physics Constraints from \nicer Pulse Profiles}


\author[0000-0002-3531-9842]{Tolga G\"uver }
\affiliation{Istanbul University, Science Faculty, Department of Astronomy and Space Sciences, Beyaz\i t, 34119, \.Istanbul, T\"urkiye}
\affiliation{Istanbul University Observatory Research and Application Center, Istanbul University 34119, \.Istanbul, T\"urkiye}
\affiliation{School of Physics, Georgia Institute of Technology, 837 State Street NW, Atlanta, GA 30332, USA}
\email{tolga.guver@istanbul.edu.tr}
\author[0000-0003-1035-3240]{Dimitrios Psaltis}
\affiliation{School of Physics, Georgia Institute of Technology, 837 State Street NW, Atlanta, GA 30332, USA}
\author[0000-0003-4413-1523]{Feryal \"Ozel}
\affiliation{School of Physics, Georgia Institute of Technology, 837 State Street NW, Atlanta, GA 30332, USA}
\author[0009-0003-6348-7143]{Tong Zhao}
\affiliation{School of Physics, Georgia Institute of Technology, 837 State Street NW, Atlanta, GA 30332, USA}

\begin{abstract}
     Modeling energy-dependent X-ray pulse profiles from rotation-powered millisecond pulsars observed with \textit{NICER} has emerged as a promising avenue for measuring neutron star radii and probing the equation of state of cold, ultra-dense matter. However, pulse profile models have often required an unwieldy number of parameters to account for complex surface emission geometries, introducing the risk of overfitting and degeneracies. To explore the number of model parameters that can be inferred uniquely, we perform a quantitative assessment of the information content in X-ray pulse profiles by applying Fourier methods. We determine the number of independent observables that can be reliably extracted from the pulse shapes, as well as from complementary X-ray spectral data obtained with \textit{XMM-Newton}, for key \textit{NICER} targets. Our analysis provides a framework for evaluating the match between model complexity and data constraints. It also demonstrates the importance of incorporating in the model the pulsed components of the magnetospheric non-thermal emission, which may often contribute significantly to the observed spectra. Our results highlight limitations in previous inferences of neutron-star radii from \nicer observations, which may have incorporated model complexity not supported by the data.
\end{abstract}
\keywords{Millisecond pulsars; X-ray astronomy; Neutron stars}

\section{Introduction}

Neutron stars have been one of the key astrophysical sources for understanding the nature of cold matter at or above nuclear saturation density \citep[see, e.g.,][]{2001ApJ...550..426L,2016ARA&A..54..401O}.  Because the Tolman-Oppenheimer-Volkoff equations allow us to relate the macroscopic measurables, namely the masses and radii, to the details of the equation of state of matter inside neutron stars~\citep{PhysRev.55.364,PhysRev.55.374}, astrophysical observations of these objects have played a central role in constraining dense matter physics. Neutron stars have especially been observed in radio, X-rays, and gamma-ray bands. Although the radio and gamma-ray emission of neutron stars generally emanate not from the surface but rather from the magnetosphere, exploiting the general relativistic effects in the timing observations of binary millisecond pulsars has proven to be a fruitful and precise tool for determining the masses of neutron stars in binary systems \citep[see e.g.][]{2008LRR....11....8L,2016ARA&A..54..401O}. However, obtaining independent measurements of radii has been significantly more challenging. 

One approach has involved X-ray observations of thermonuclear X-ray bursts \citep[][]{1993SSRv...62..223L} that reach the Eddington luminosity. In such cases, the X-ray flux measured at or around the time of the observed contraction of the photosphere is related to the mass and radius of the neutron star \cite[][]{paradijs1978average-3e6,paradijs1978interpretation-d1c,1990A&A...237..103D,paradijs1993constraints-b0c}. Furthermore, during the cooling tails of these events, there is again a relationship between 
the apparent emitting area and the mass and the radius of the neutron star \citep{1982A&A...107...51V,paradijs1993constraints-b0c}. Observations of bursting neutron stars with the Proportional Counter Array (PCA) onboard the Rossi X-ray Timing Explorer (RXTE) employing this technique resulted in multiple radius measurements in the $10-11.5\;$km range \citep{zel2009mass-0a5,2010ApJ...712..964G,2010ApJ...719.1807G,2010PhRvD..82j1301O,Steiner2010,Steiner2013,2016ApJ...820...28O}.

Another approach has utilized imaging X-ray telescopes with very low backgrounds, which allowed studies of the surface thermal X-ray emission from neutron stars to constrain their masses and radii. Specifically, observations of low mass X-ray binaries in quiescence in globular clusters have yielded correlated constraints that also favor smaller radii \citep[$\lesssim$ 11.5~km, see, e.g.,][]{2014MNRAS.444..443H,2016ApJ...831..184B,2018MNRAS.476.4713S,2020MNRAS.495.4508E}.

A third avenue toward constraining neutron star masses and radii is through modeling the energy-dependent X-ray pulse profiles \citep[see, e.g.,][]{pechenick1983hot-192,weinberg2001oscillation-e8a, 2014ApJ...792...87P,2016RvMP...88b1001W}. Such measurements require sensitivity to soft X-rays, large effective area, and very high time resolution, all of which are provided by the Neutron Star Interior and Composition Explorer (NICER, \citealt{2012SPIE.8443E..13G}. Long \nicer observations of \src, \psrdz, \psrtt, \psrzs allowed for their X-ray pulse profiles to be measured sensitively. Subsequent studies modeled the shapes, locations, and numbers of hot spots on the surfaces of these pulsars and, based on these models, derived constraints on their masses and radii \citep{miller2019psr-ae0,dittmann2024more-4c2,riley2019nicer-f21,riley2021nicer-590,salmi2024nicer-bfc,salmi2024radius-449,Choudhury_2024}. In the majority of these studies, complex and often unexpected geometries were inferred, leading to a proliferation of model parameters to describe the shapes and locations of multiple hot spots on the stellar surface. 

The ability of an observed pulse profile to uniquely constrain the surface emission characteristics and, therefore, the stellar mass and radius relies heavily on the complexity observed in the pulse profile and the mapping of its features to the surface emission. For example, if a pulse profile is well described by a simple sinusoid (within the measurement uncertainties), it provides at most two pieces of information: amplitude and phase. Such limited information may be insufficient to constrain a model with a complex emission structure and will lead to overfitting and degeneracies. This raises a fundamental question: what is the information content of an observed pulse profile, and how does it compare to the number of parameters used to model it?

In this paper, we use Fourier methods to determine the number of model parameters that can be measured from a combination of the pulse profiles observed with \nicer\ and the X-ray spectral data observed with \xmm for the primary \nicer targets. In \S\ref{sec:method} we outline the methodology, in \S\ref{sec:pulse_profiles}-\ref{sec:merging} we present our analysis of the data, and in \S\ref{sec:concl}, we discuss our findings.

\section{Methodology}
\label{sec:method}

Pulsed emission from millisecond pulsars in the X-rays is most likely caused by returning magnetospheric relativistic charged particles causing temperature anisotropies on the surface \citep[see e.g.,][]{1975ApJ...196...51R,Philipov2015,Brambilla2018,baubck2019atmospheric-933} as well as non-thermal magnetospheric processes. Although the detailed nature and spin-phase dependence of the non-thermal emission is not clear, the observational properties of the thermal component(s) have been studied extensively \citep[see e.g.,][]{2002nsps.conf..263Z,Bogdanov2007,Bogdanov2008,Bogdanov2009,2013RPPh...76a6901O,Bogdanov2013}. 

Reflecting mainly the magnetic field configuration and the properties of the atmosphere, heating anisotropies caused by return currents can create hot spots with different sizes and temperatures at different locations on the surface, which then form the observed pulsed emission as the neutron star rotates. The shape of the pulse profile depends on the number, size, temperature, and relative orientation of the different hot spots. Each one contributes a distinct component to the X-ray spectrum and introduces a unique set of Fourier modes to the photon flux as a function of spin phase.  Furthermore, relativistic effects caused by the strong gravitational field of the neutron star reshape this pulsed emission, resulting in the observed pulse profiles \citep[][]{1983ApJ...274..846P,poutanen2006pulse-c08, 2011ApJ...726...56M,2014ApJ...792...87P}.

We write the contribution of the $i-$th surface component to the flux at energy $E$ and spin phase $\phi$ as
\begin{eqnarray}
    F_i(E,\phi)&=&\sum_{n=0}^\infty \left\{A_{in}(E)\cos\left[n(\phi-\phi_i)\right]\right.\nonumber\\
    &&\qquad\qquad\left.+B_{in} (E)\sin\left[n(\phi-\phi_i)\right]\right\}
\end{eqnarray}
where $A_{in}$ and $B_{in}$ are the amplitudes of the cosine and sine Fourier modes and $\phi_i$ is the central azimuth of the hot spot that generates this spectral component with respect to some fiducial value. For slowly spinning neutron stars \citep[<300~Hz, see][]{2014ApJ...792...87P}, the sine terms in this expansion are subdominant (see, e.g., \citealt{poutanen2006pulse-c08}) and, for brevity, we will not carry them hereafter. The total flux from all $S$ surface components is then
\begin{equation}
    F(E,\phi)=\sum_{i=1}^S\sum_{n=0}^\infty A_{in}(E)\cos\left[n(\phi-\phi_i)\right]\;.
\end{equation}
Expanding the cosine terms, we can write this as
\begin{eqnarray}
     F(E,\phi)&=&\sum_{n=0}^\infty\sum_{i=1}^S \left[A_{in}(E)\cos(n\phi_i)\cos(n\phi)\right.\nonumber\\
     &&\qquad\qquad\left.+A_{in}(E)\sin(n\phi_i)\sin(n\phi)\right]
\end{eqnarray}
or equivalently
\begin{equation}
    F(E,\phi) = A_0(E) + \sum_{n=1}^\infty [A_{n}(E)\cos{(n\phi)}+B_n(E)\sin{(n\phi)}] 
\label{eq:expansion}\;,
\end{equation}
where 
\begin{eqnarray}
    A_0(E)&=&\sum_{i=1}^S A_{i0}(E)\label{eq:A0}\\
    A_n(E)&=&\sum_{i=1}^S A_{in}(E)\cos(n\phi_i)\\
    B_n(E)&=&\sum_{i=1}^S A_{in}(E)\sin(n\phi_i)\;.\label{eq:Bn}
\end{eqnarray}

Equation~(\ref{eq:expansion}) is a complete expression that can describe any pulse profile of arbitrary complexity. In practice, however, an infinite number of Fourier components are not required to describe or supported by the actual data, both because of the signal-to-noise of the data and of the degree of the intrinsic complexity of the pulse profiles. As a result, the sum of Fourier modes in Eq.~(\ref{eq:expansion}) terminates at some $N=N_{\rm max}$. In \S3, we use \nicer data to quantify the number of independent Fourier components that can be measured at different photon energies for the primary targets. In \S4, we use \xmm data to determine the number $S$ of independent spectral components arising from the stellar surface. We combine the results of these two analyses in \S5 to quantify the number of  independent model parameters that can be meaningfully inferred from the current \nicer data. We note that, even if the observations provide only upper limits on the 
coefficients of additional Fourier modes, these upper limits on their magnitudes can still be used to exclude regions of the model parameter space. In other words, upper limits on higher-order Fourier modes can still rule out models that predict large amplitudes. However, this can only serve to put bounds on the physical parameters of interest (e.g, masses and radii of neutron stars) and perhaps exclude particular equations of state but do not lead to actual measurements.

We focus our attention to \src given the fact that it is the brightest among all the pulsars used for pulse profile modeling \citep{2016MNRAS.463.2612G}. In Appendix~\ref{app:others}, we repeat the same analysis for the remaining sources \psrdz, \psrzs, \psrtt, which have been observed for pulse profile modeling with \nicer. 

\src is the closest pulsar with a known distance of 156.96$\pm$0.11~pc \citep{2024ApJ...971L..18R}. It was discovered during the Parkes radio pulsar survey \citep{johnston1993discovery-e63} and as an X-ray source with ROSAT \citep{1993Natur.365..528B}. Being the closest, brightest millisecond pulsar with a precise mass measurement (M=1.418$\pm$0.044 M$_{\sun}$, \citealt{2024ApJ...971L..18R}) made \src one of the prime targets for \nicer. 

Its broadband X-ray spectrum has also been studied extensively. Using data from ROSAT, \xmm, and \nustar, \cite{2016MNRAS.463.2612G} provided detailed constraints on the number and parameters of the X-ray spectral components. They found that the 0.1--20~keV band data are best described with three thermal components and one power-law component, and determined the amount of absorption by the interstellar medium. Most importantly, thanks to \nustar observations, they detected pulsed emission from \src with a pulse fraction of 24$\pm$6\% in the 2–20 keV band. \cite{2016MNRAS.463.2612G} also demonstrated that the contribution from the power-law component produces comparable flux to that of the surface emission at energies above 2~keV, although the exact amount changes depending on the selected model for the thermal emission components.

\section{NICER Data and Pulse Profiles}
\label{sec:pulse_profiles}
We obtained the NICER data extracted by \cite{Choudhury_2024} from the Zenodo archive\footnote{https://zenodo.org/records/13766753}. Specifically we use the file, \emph{j0437\_3c50\_cl\_evt\_merged\_phase.txt}, which contains time of arrivals (ToAs) of each event as well as their PI (energies) and the spin phase values, such that an energy dependent pulse profile can be easily generated for the data obtained with \nicer from 2017 July 6 to 2021 October 11. For the background, we used the \emph{rj0437\_3c50\_bg\_pha\_spec.txt} file. This file contains channels, count rates, and the statistical errors, as calculated by the 3C50 model. In addition to this background, there is a contribution to the number of detected photons from a nearby AGN (RX~J0437.4$-$4711) in the \nicer field of view \citep{Choudhury_2024}. The AGN counts are calculated and provided as a separate spectral file, \emph{rscale42\_rxj0437\_ep.pha}, in the Zenodo archive. We use this file to take into account the AGN contribution.

\begin{figure*}
    \centering
    \includegraphics[scale=0.35]{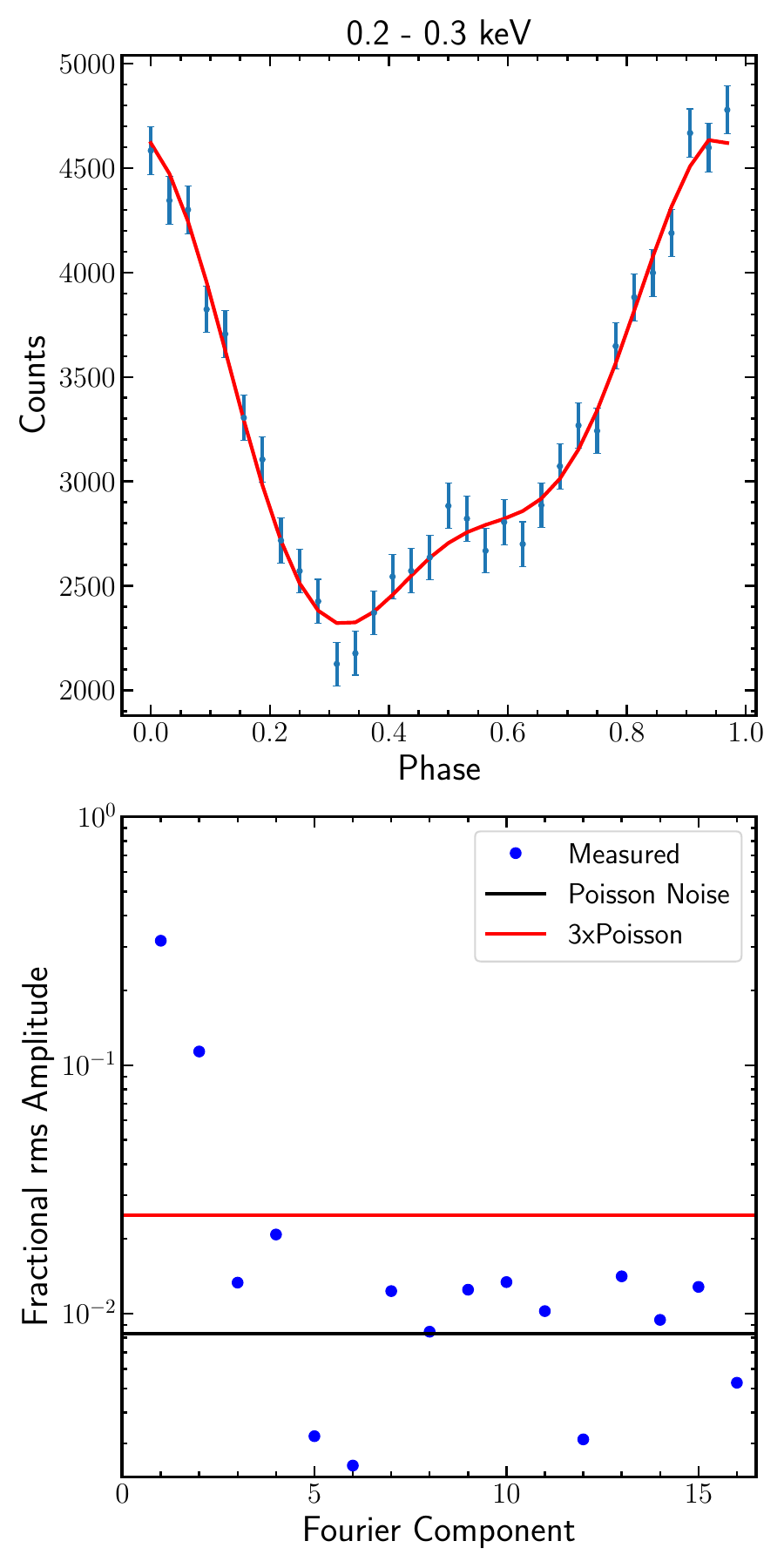}
        \includegraphics[scale=0.35]{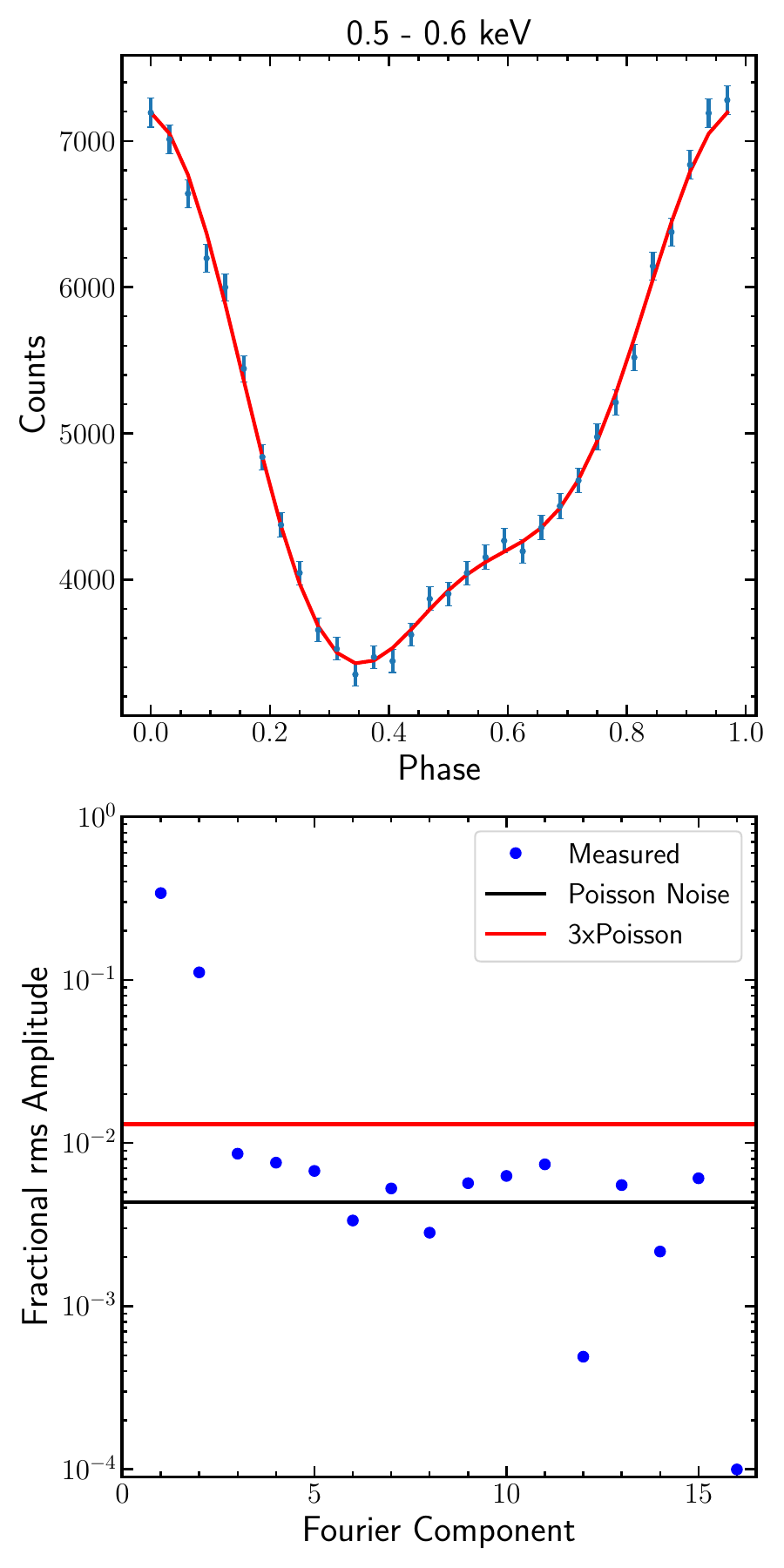}
            \includegraphics[scale=0.35]{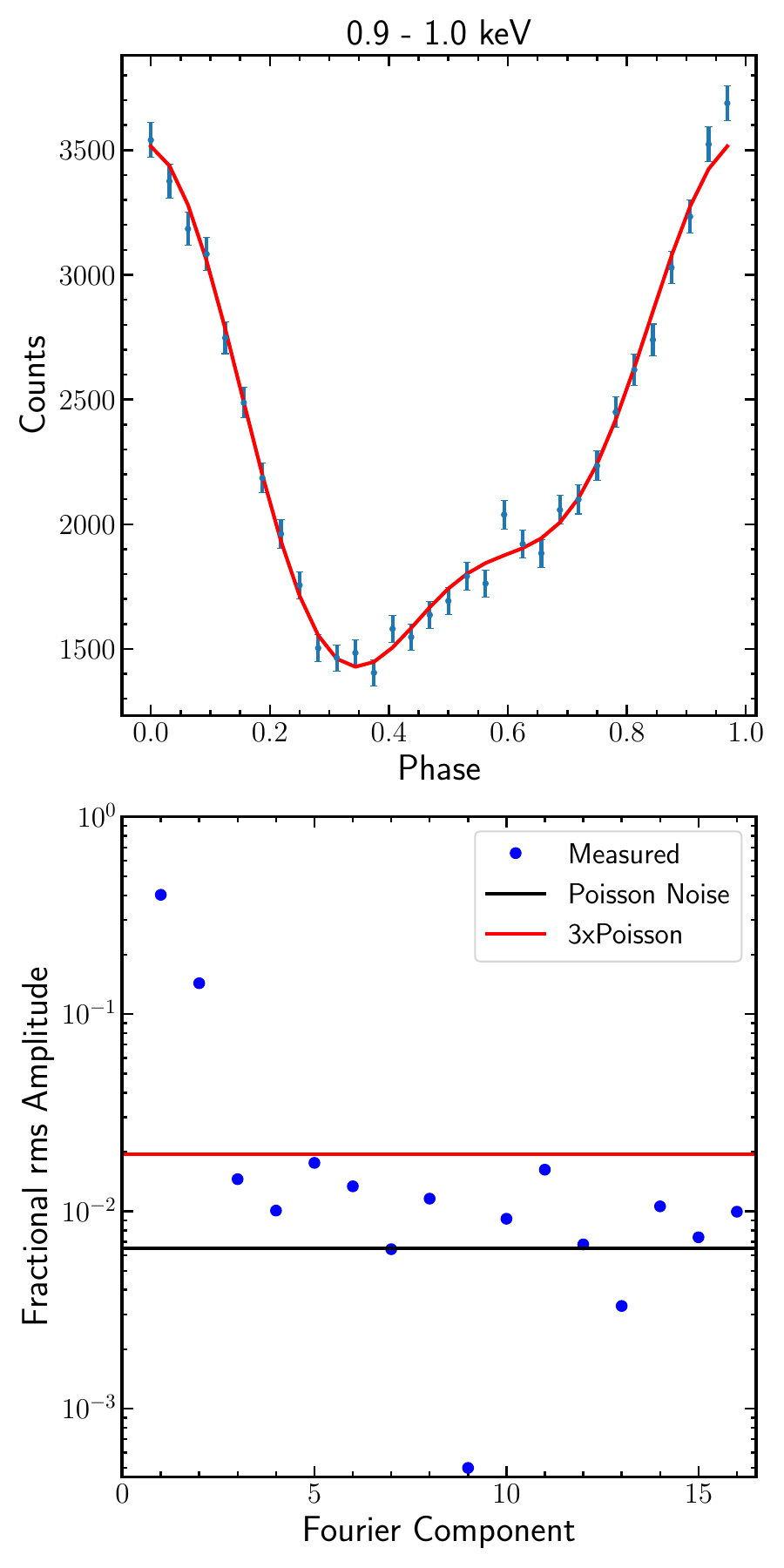}

    \includegraphics[scale=0.35]{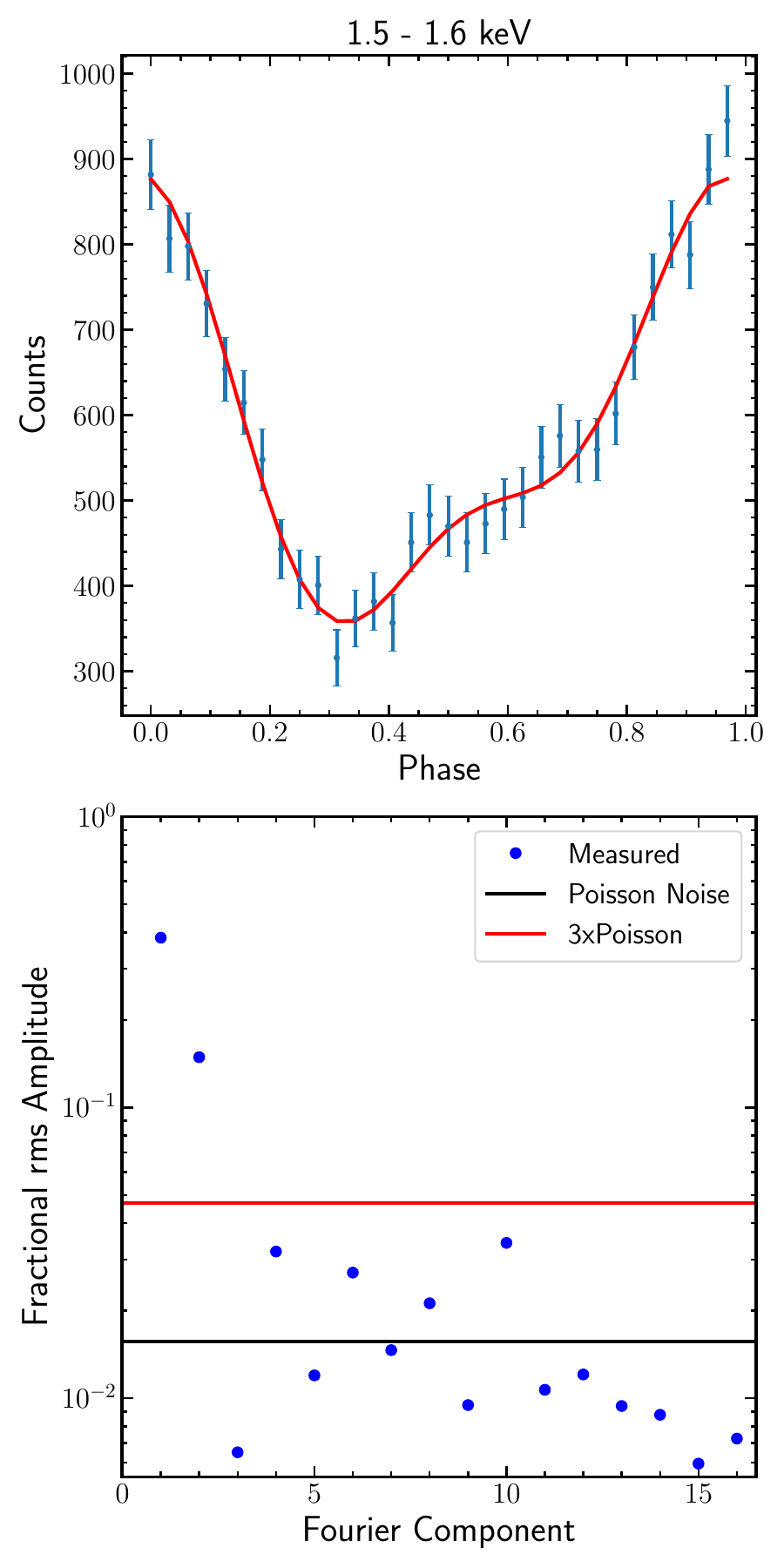}
        \includegraphics[scale=0.35]{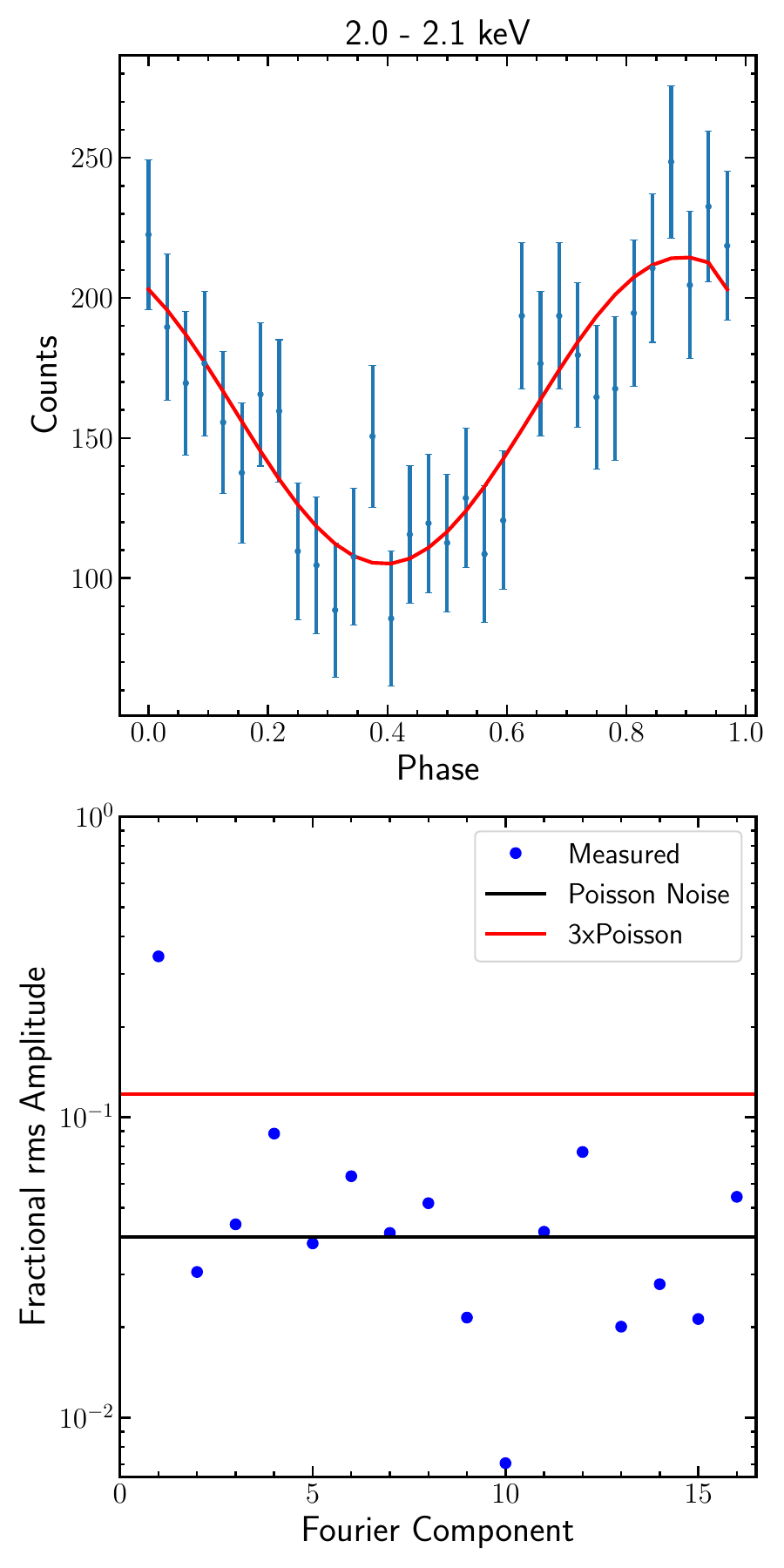}
            \includegraphics[scale=0.35]{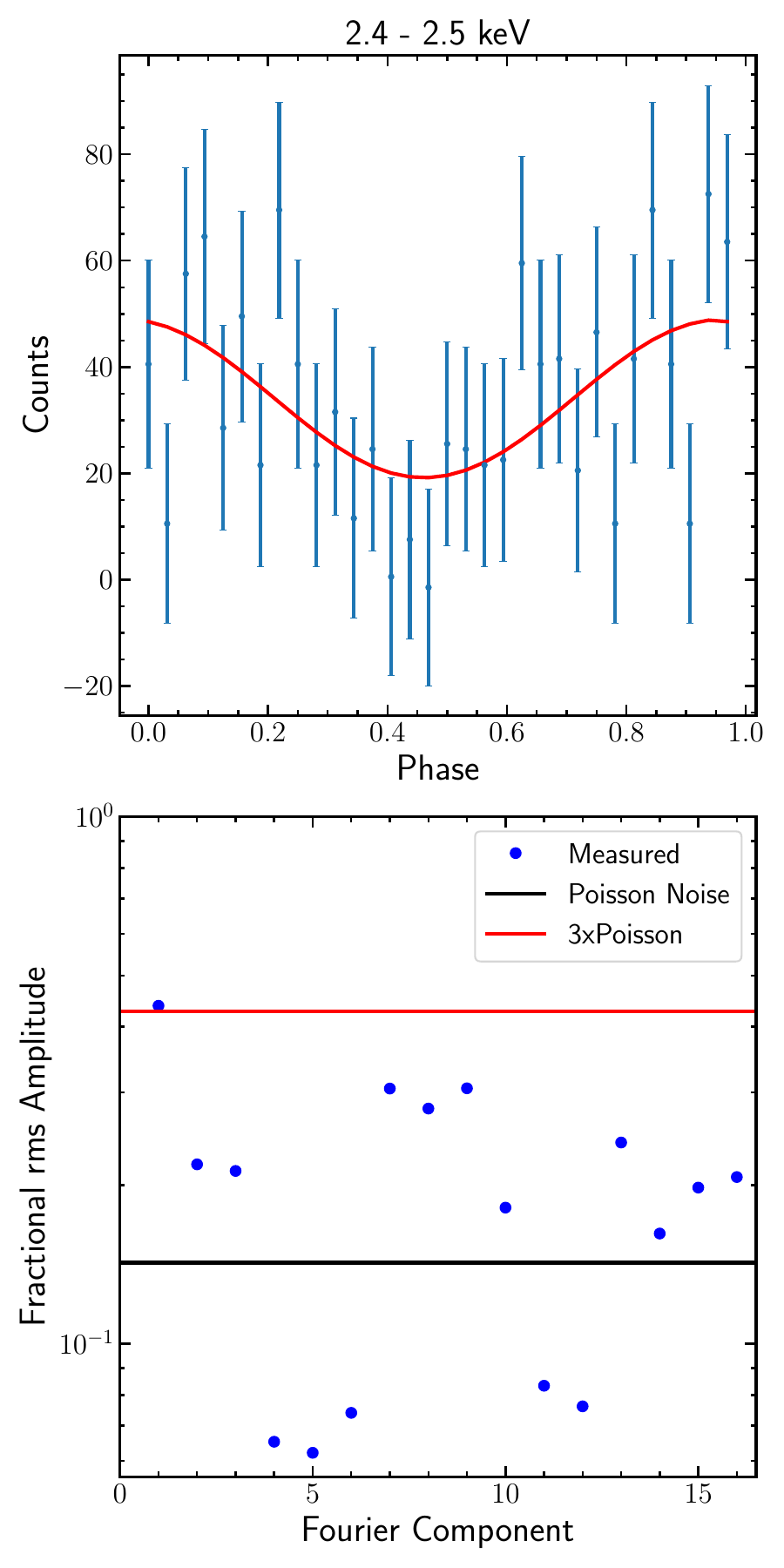}
     
    \caption{Pulse profiles (first and third rows) and measured amplitudes of the Fourier components (second and fourth rows) in a sample of energy bands for \src. The horizontal black lines show the expected amplitude of the Poisson noise, while the red lines show our detection threshold of three times the Poisson noise. At most two Fourier components are measurable above the noise in each of the energy bands.}
    \label{fig:pp1}
\end{figure*}

Since the full event file contains the total number of events arriving at the detectors throughout the numerous observations, it is necessary to calculate the total number of background events in each energy range. For that purpose, we used the total exposure time of 1.328~Ms \citep{Choudhury_2024}. Within the 0.3--3.0~keV range, the event file has a total of 2,095,698 events. In the same energy range, the 3C50 background model predicts 0.495 counts~s$^{-1}$, which results in a total of 657,677 background events. When calculating the pulse profiles, we divided the $0.3-3.0$~keV range into 0.1~keV energy bands. In \autoref{tab:charac}, we provide the total number of source and background events for each energy band. Using these net counts, we generated pulse profiles, each containing 32 phases. Several example pulse profiles are shown in \autoref{fig:pp1}. 


\begin{table}[t!]
\centering
\caption{Observed Counts and Backgrounds for \src}
\begin{tabular}{cccc}
\hline 
Energy Range & Tot. Events$^{1}$& Bkg. Counts$^{2}$& Net Counts$^{1}$\\
(keV) &\\
\hline
0.3--0.4 & 303958 & 97695 & 206263 \\
0.4--0.5 & 287740 & 89671 & 198069 \\
0.5--0.6 & 243134 & 82817 & 160317 \\
0.6--0.7 & 197956 & 64474 & 133482 \\
0.7--0.8 & 165169 & 53197 & 111972 \\
0.8--0.9 & 138230 & 46702 & 91528 \\
0.9--1.0 & 114619 & 40919 & 73700 \\
1.0--1.1 & 104235 & 39161 & 65074 \\
1.1--1.2 & 71086 & 28381 & 42705 \\
1.2--1.3 & 67159 & 29166 & 37993 \\
1.3--1.4 & 57892 & 26864 & 31028 \\
1.4--1.5 & 51396 & 26652 & 24744 \\
1.5--1.6 & 42896 & 24165 & 18731 \\
1.6--1.7 & 36301 & 21649 & 14652 \\
1.7--1.8 & 33078 & 21099 & 11979 \\
1.8--1.9 & 27596 & 18710 & 8886 \\
1.9--2.0 & 23091 & 16733 & 6358 \\
2.0--2.1 & 20775 & 15661 & 5114 \\
2.1--2.2 & 20049 & 16086 & 3963 \\
2.2--2.3 & 13169 & 11280 & 1889 \\
2.3--2.4 & 12557 & 11487 & 1070 \\
2.4--2.5 & 12081 & 10991 & 1090 \\
2.5--2.6 & 11474 & 10291 & 1183 \\
2.6--2.7 & 10778 & 10284 & 494 \\
2.7--2.8 & 10033 & 9535 & 498 \\
2.8--2.9 & 9731 & 9434 & 297 \\
2.9--3.0 & 9515 & 9077 & 438 \\
\hline
\end{tabular}
\label{tab:charac}\\
\footnotesize{$^{1}$ Total number of events within each energy band.\\}
\footnotesize{$^{2}$ Rounded total number of background events including the contribution from the AGN within each energy band.}
\end{table}

We then calculated the discrete Fourier transforms (FFTs) of each pulse profile, using the \emph{numpy.fft} algorithm \citep{harris2020array}. We evaluated the amplitudes of each Fourier component as $(A_n^2+B_n^2)^{1/2}$ and of the Poisson noise using equation~(A6) of \cite{psaltis2014}. In Figure~\ref{fig:pp1}, we show the measured Fourier amplitudes and the Poisson noise for a selected group of energy bands (see also Appendix~\ref{app:det_fig} for details). We adopt a measurement threshold of three times the Poisson noise for each energy band. Given that we perform Fourier analysis in 17 energy bands, for each pulsar, and measure 16 Fourier components in each band, this $3\sigma$ threshold corresponds to at most $\sim (1-0.997)\times 17\times 16=0.8<1$ spurious detections. We count the number of Fourier components with amplitudes larger than this threshold find that at most 2 components can be measured in each energy band.

\begin{figure}
    \centering
            \includegraphics[width=\columnwidth]{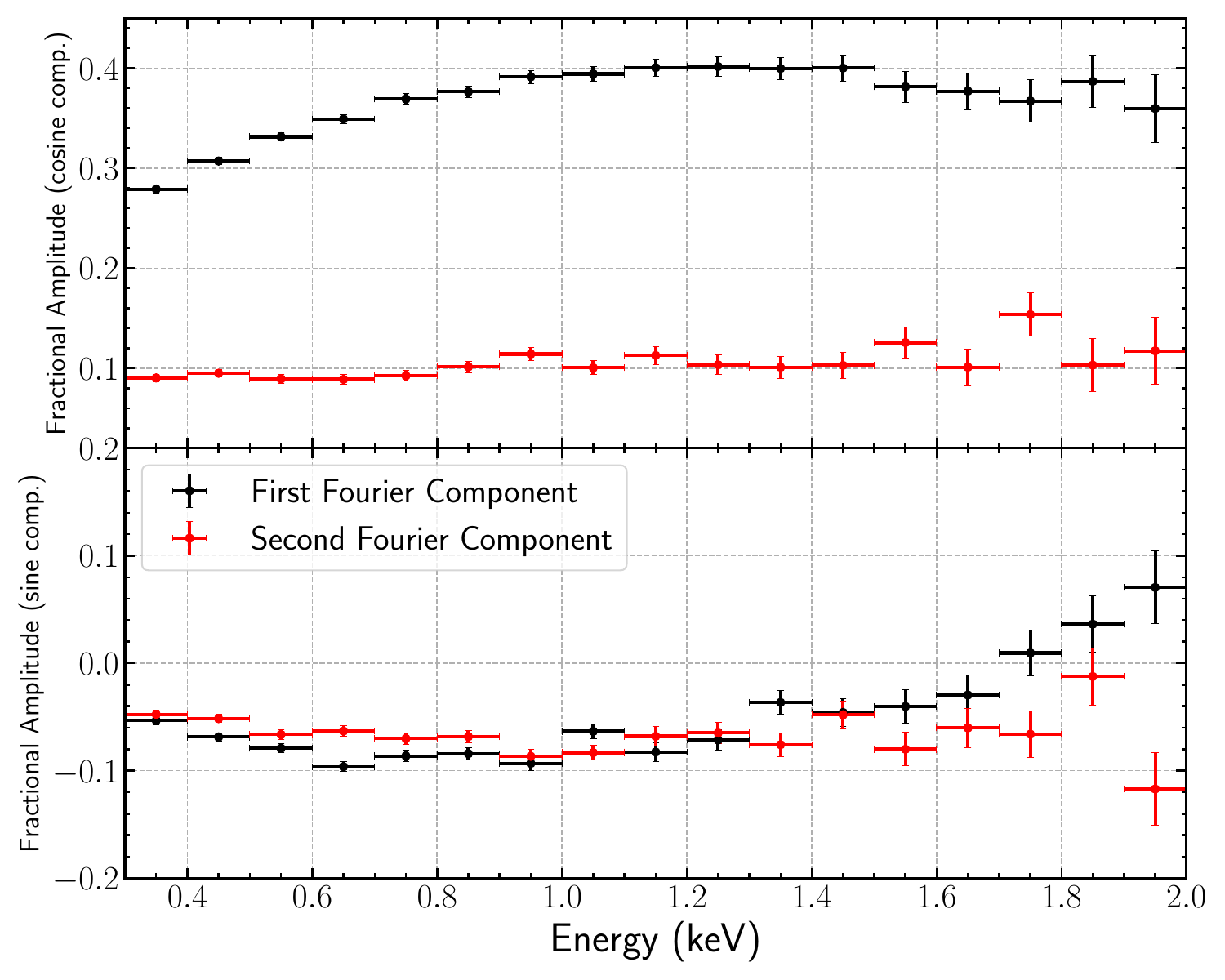}
            \caption{Amplitudes of the first two cosine (upper panel) and sine (lower panel) Fourier components as a function of energy for \src. The weak dependence  of these amplitudes on energy significantly reduces the number of independent model parameters that can be inferred from pulse profile modeling.}
    \label{fig:fourier_comp}
\end{figure}

Figure~\ref{fig:fourier_comp} shows the individual amplitudes of the cosine (top panel) and sine (bottom panel) Fourier components as a function of photon energy, i.e.,  $A_n(E)/A_0$ and $B_n(E)/B_0$, for $n=1,2$. We see a weak energy dependence in the amplitude of the first cosine Fourier component and negligible ones in all the other components up to $\sim 1.7$~keV. This weak dependence significantly reduces the number of independent model parameters one can infer using the energy dependence of the pulse amplitudes, as we will discuss in \S5. In addition, it is notable that above $\sim 1.7$~ keV, the first sine component changes sign. We will show in \S4 that this is related to the increasing contribution of the magnetospheric power-law component around this energy. 

\section{X-ray Spectral Analysis}
\label{sec:sp}

In this section, we perform a spectral analysis of \src using the longest \xmm observations. Our main goal with this analysis is to {\em (i)\/} obtain flux ratios of the model components as a function of energy and {\em (ii)\/} show the sensitivity of these flux ratios on the assumptions and parameter uncertainties of the spectral model. 

For this analysis, we  used the \xmm observation of \src (OBSID: 0603460101) and extracted MOS1 and MOS2 spectra. Note that, during the same observation, EPIC-pn data were obtained in timing mode. However, because the high background of this mode prevents any meaningful spectral analysis, we neglected this data set. For calibration, we used \emph{emproc} within XMMSAS, version 20230412\_1735-21.0.0, and the latest available calibration files as of January 2025. We used circular regions centered around the source to extract the source spectrum, and nearby similar-sized regions to extract background spectra. We also used the \emph{rmfgen} and \emph{arfgen} tools to generate response and ancillary response matrix files, respectively. Finally, we grouped the energy bins to contain at least 150 counts per energy channel.

\begin{figure}
    \centering
    \includegraphics[width=1.1\columnwidth]{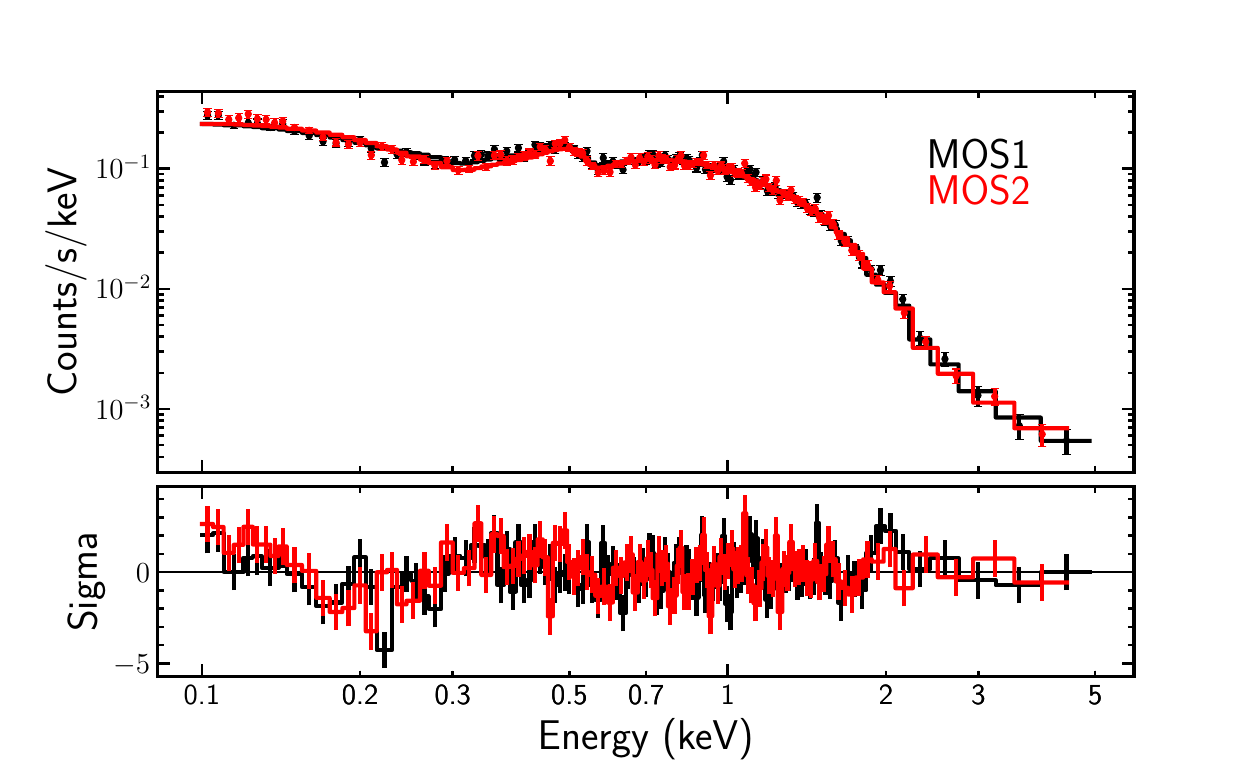}
    \caption{X-ray spectra  of \src, obtained with \xmm MOS1 (black) and MOS2 (red) together with the best fit model that contains three blackbodies plus a power-law component, all absorbed due to ISM. The model parameters are from a simultaneous fit  to ROSAT, \xmm and \nustar data  \citep{2016MNRAS.463.2612G}.}
    \label{fig:xmm_fit}
\end{figure}


Following \cite{2016MNRAS.463.2612G}, we fit the X-ray spectrum of \src using a combination of three absorbed blackbodies plus a power-law component. Since the \xmm data only allow us to fit the $0.1 -5.0$~keV range, we fix the photon index of the power-law component and its normalization to the best-fit values obtained from the analysis of the \nustar data by \cite{2016MNRAS.463.2612G} and use $\Gamma = 1.65\pm 0.24$ and a normalization of (2.1$_{-0.7}^{+0.9})\times10^{-5}$ photons/keV/cm$^2$/s at 1~keV. In \autoref{fig:xmm_fit} we show the MOS1 and MOS2 data, the best-fit spectral model with parameters fixed to the values given in \cite{2016MNRAS.463.2612G}, and the residuals in \autoref{fig:xmm_fit}. We note that the X-ray data alone cannot put strong constraints on the hydrogen column density to \src, which plays a significant role in determining the relative contributions of the surface components below $\sim 0.4$~keV. 

Numerous studies have reported N$_{\rm H}$ values for this pulsar, showing modest variations among them. 
For example, \cite{2024ApJ...971L..18R} find $ 0.8 \times 10^{20}~\rm{cm}^{-2}$, while the HEASARC NH Tool\footnote{https://heasarc.gsfc.nasa.gov/cgi-bin/Tools/w3nh/w3nh.pl} gives $0.2 \times 10^{20}~\rm{cm}^{-2}$ based on neutral Hydrogen observations \citep{2016A&A...594A.116H}. \cite{2019MNRAS.490.5848G} report $0.7 - 2.4 \times 10^{20}~\rm{cm}^{-2}$ using UV and soft X-ray data. Finally, \cite{2016MNRAS.463.2612G} infer $N_{\rm H} = 2.4 \times 10^{20}~\rm{cm}^{-2}$ from their joint fit to \xmm, ROSAT, and \nustar data. In contrast, \cite{Choudhury_2024} report a 68\% credible interval of $(0.05_{-0.04}^{+0.68}) \times 10^{20}~\rm{cm}^{-2}$ based on \nicer observations, which includes values that are an order of magnitude smaller than any of the previously reported values and the large uncertainties show that the parameter is unconstrained.

\begin{figure}
    \centering
    \includegraphics[scale=0.4]{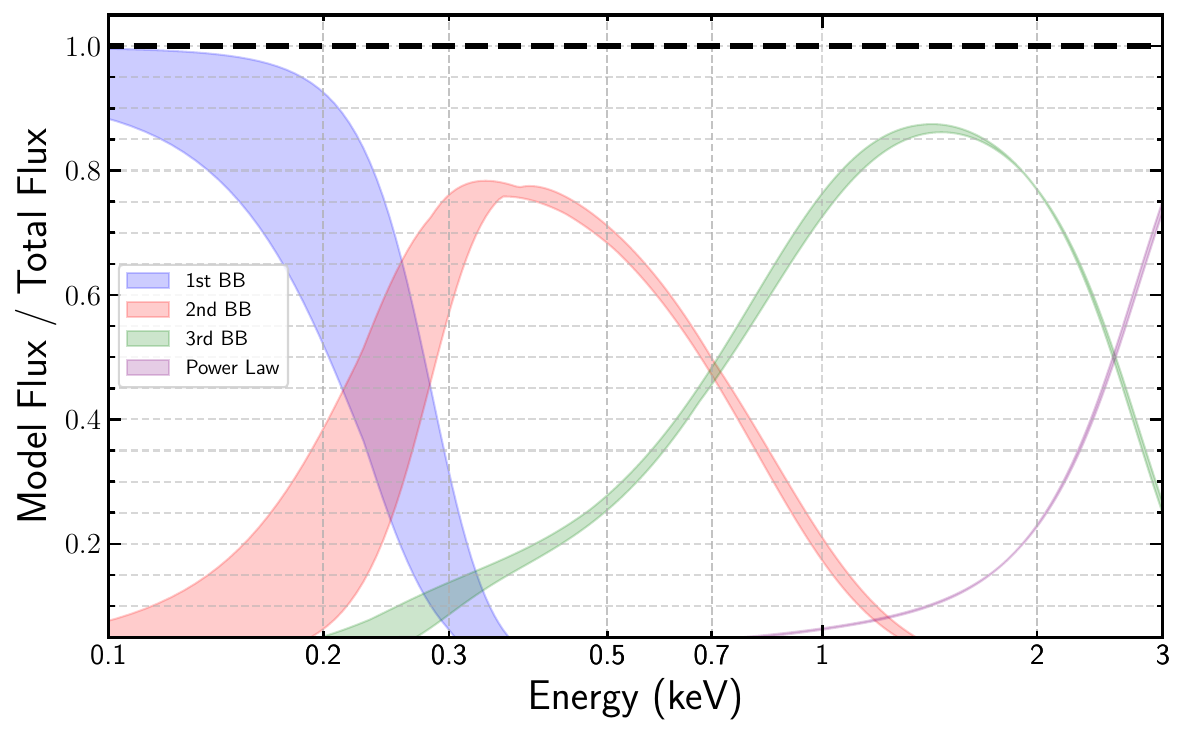}

    \caption{Fractional contribution to the total flux of each model component as a function of energy. The bands correspond to varying the hydrogen column density within the range reported in the literature.}
    \label{fig:sp_comp}
\end{figure}

To show the effect of the hydrogen column density value on the inferred spectral parameters, we fixed it to a sample of values covering the full range reported above, and kept the parameters of the power-law component fixed. As expected, using different values for $N_{\rm H}$ resulted in significant changes in the temperatures and normalizations of the thermal components.  In \autoref{fig:sp_comp}, we show the relative contributions of the individual spectral components to the total flux as a function of energy, with the band corresponding to the range of $N_{\rm H}$ values. In the same plot, we also show the best-fit model of \cite{2016MNRAS.463.2612G}. It is clear that the relative contributions of the coolest two blackbody components depend strongly on the assumed $N_{\rm H}$ value and vary by approximately 10\% at energies less than $\sim 1$~keV and by as much as 50\% at energies less than $\sim 0.3$~keV. 

\section{Parameter Inference from Combined Fourier and Spectral Analyses}
\label{sec:merging}

In \S3, we determined that \nicer data of \src support the measurement of at most two Fourier components in each photon energy band, i.e., $N_{\rm max}=2$. In \S4, we presented that the X-ray data from \xmm also require 3 surface emission components (blackbodies) and a magnetospheric power-law component, i.e., $S=4$. In this section, we analyze the energy dependence of the Fourier amplitudes that correspond to each spectral component in order to quantify the total number of independent parameters that can be inferred from the combined Fourier and spectral information. This will inform the maximum complexity of any model of the surface emission that can be robustly constrained by the NICER data.

We first write the energy dependence of the fractional Fourier amplitude of each spectral component as a polynomial in $E$, i.e.,
\begin{equation}
    \frac{A_{in}(E)}{A_{i0}(E)}=\left[f_{in}+f^\prime_{in} E +
    \frac{1}{2}f^{\prime\prime}_{in} E^2+...\right]\;.
\end{equation}
We then determine the number of terms in this expansion that are necessary to reproduce the observational data. We will show that each component requires only the constant term. In other words, the pulse fraction of each spectral component does not evolve as a function of energy, within the uncertainties of the current measurements.

To demonstrate this, we rewrite equations~(\ref{eq:A0})-(\ref{eq:Bn}) as
\begin{eqnarray}
    \frac{A_n(E)}{A_0(E)}&=&\sum_{i=1}^4 f_{in} \left[\frac{A_{i0}(E)}{A_0(E)}\right]\cos(n\phi_{in})
    \label{eq:fitA}\\
    \frac{B_n(E)}{A_0(E)}&=&\sum_{i=1}^4 f_{in} \left[\frac{A_{i0}(E)}{A_0(E)}\right]\sin(n\phi_{in})
    \label{eq:fitB}\;,
\end{eqnarray}
where $A_n(E)/A_0(E)$ and $B_n(E)/A_0(E)$ are the fractional cosine and sine amplitudes shown in Figure~\ref{fig:fourier_comp} and $A_{i0}(E)/A_0(E)$ are the fractional contributions to the total flux of each spectral component shown in Figure~\ref{fig:sp_comp}. Note that, for each surface component, the angular arguments $\phi_{in}$ are determined only by the azimuth of the hotspot associated with that spectral component on the neutron star surface. As such, $\phi_{in}$ should not depend on $n$. However, this is not necessarily true for the magnetospheric power-law component. When fitting the data, we will allow for these angular arguments to depend on $n$ even for the thermal components and show that they are consistent with being constant between the various Fourier components. 

  \begin{figure}
     \centering  
         \includegraphics[scale=0.4]{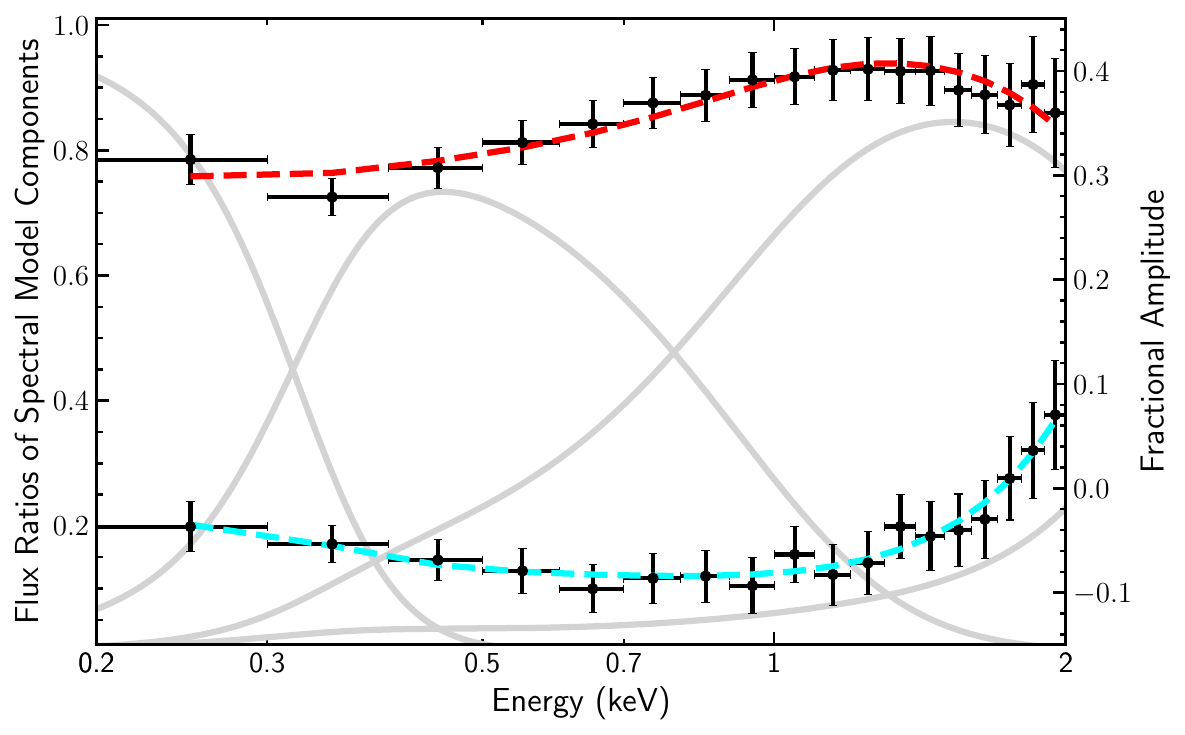}
   \includegraphics[scale=0.4]{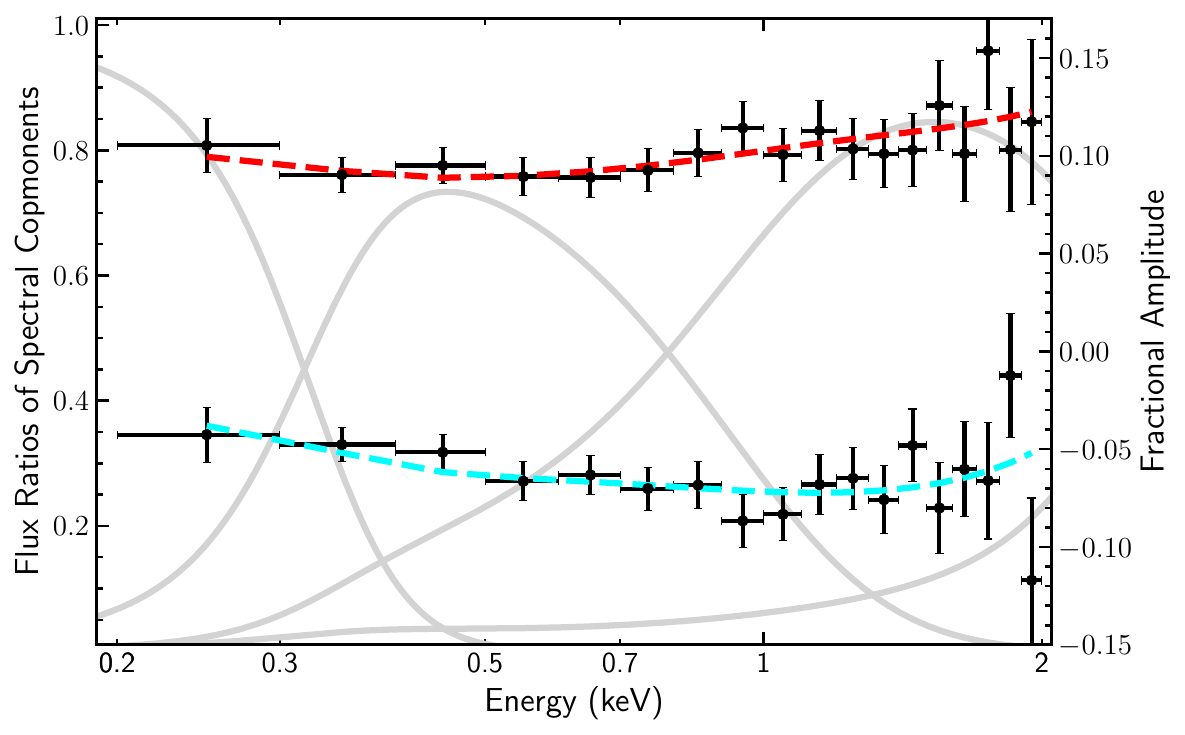}
     \caption{ Grey curves show the fractional flux contributions of the four spectral components (see Fig.~\ref{fig:sp_comp}). Error bars show the fractional amplitudes of the fundamental (left panel) and harmonic (right panel) cosine and sine components of the pulse profiles (see Fig.~\ref{fig:fourier_comp}). The dashed lines show the predicted fractional amplitudes for the model described in the text.}
     \label{fig:fr_comp_model}
  \end{figure}

We obtain the parameters $f_{in}$ and $\phi_{in}$ by fitting equations~(\ref{eq:fitA})-(\ref{eq:fitB}) to the Fourier components we measured in \S3, using the fractional flux contributions of the spectral components we inferred in \S4. We only use data up to 2~keV, beyond which the magnetospheric power-law component contributes significantly. We account for the uncertainty introduced by $N_{\rm H}$ on the fractional flux contributions of the spectral components by adding a 5\% systematic uncertainty to the measurements (see Fig.~\ref{fig:sp_comp}). 

In order to quantify the posteriors and explore the correlations between the inferred parameters, we employed a modified version of the Markov Chain Monte Carlo algorithm MARCH \citep[MArkov Chains for Horizons,][]{2022ApJ...928...55P}. Originally developed for fitting interferometric data from the Event Horizon Telescope, MARCH, uses parallel tempering and is optimized for high performance. In the calculation, we assumed flat priors in amplitude from 0-1 and in phase from 0-360 degrees.

\begin{figure*}
    \centering

   \includegraphics[scale=0.4]{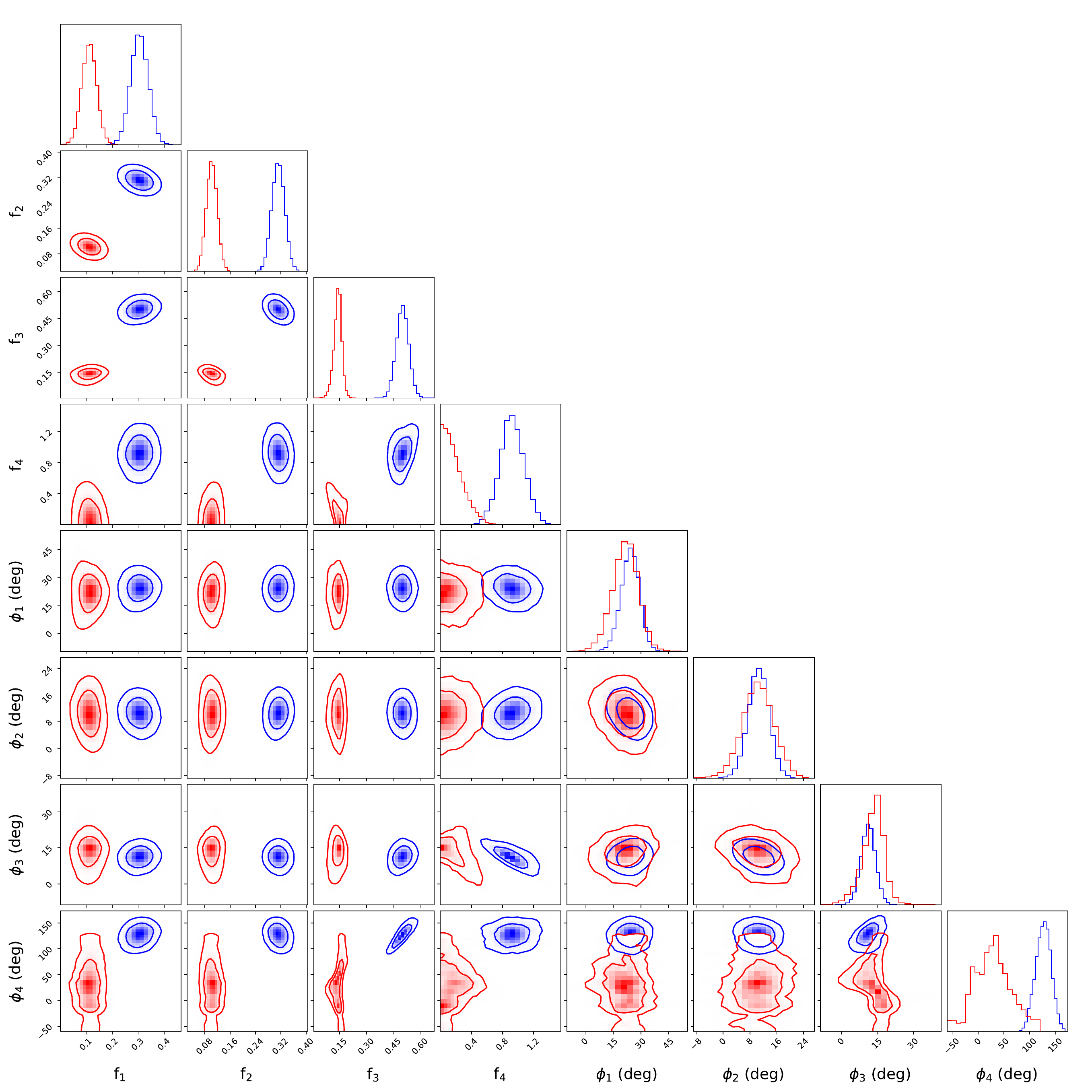}
     \caption{ The posteriors of the fractional Fourier amplitudes $f_{in}$ and angular arguments $\phi_{in}$ of the four spectral components in \src. Blue curves correspond to the fundamental components ($n=1$) and red curves to the harmonics ($n=2$).}
     \label{fig:corner}
 \end{figure*}

\begin{table*}[ht!]
\centering
\caption{Spectral components and their corresponding contributions to Fourier amplitudes}
\begin{tabular}{cccccc}
\hline
&&\multicolumn{2}{c}{Fundamental} & \multicolumn{2}{c}{Harmonic}\\
Spectral Component & Parameters$^{1}$ & Amplitude& Phase & Amplitude & Phase \\
 & kT (eV), R (km)& $f_{i1}$ & $\phi_{i1}$ (deg)& $f_{i2}$ & $\phi_{i2}$ (deg)\\
\hline
Blackbody 1 & 32, 10.98  & 0.30 $\pm$0.03 & 24.01$^{+4.86}_{-4.99}$ &0.11$\pm$0.03 &21.12$^{+6.58}_{-7.16}$\\
Blackbody 2 & 125, 0.25 &0.31 $\pm$0.02 &11.07$^{+3.08}_{-3.12}$ & 0.10$\pm$0.02&10.57$^{+4.60}_{-4.53}$\\
Blackbody 3 & 271, 0.04 & 0.51 $\pm$0.03&11.11$^{+2.92}_{-3.03}$ &0.14$\pm$0.02 &9.69$^{+3.57}_{-4.99}$\\
\hline
Power-law & $\Gamma$=1.65, 2.1$\times10^{-5}$ & 0.92 $\pm$0.16&128.32$^{+14.27}_{-15.53}$ & 0.209$^{+0.17}_{-0.11}$&42.33$^{+36.24}_{-35.81}$\\
\hline
\end{tabular}
\label{tab:res}\\
\footnotesize{$^{1}$ Values are adopted from \citet{2016MNRAS.463.2612G}.}
\end{table*}

\autoref{fig:fr_comp_model} combines the information from Figures~\ref{fig:fourier_comp} and \ref{fig:sp_comp} and also shows the inferred best-fit model (eqs.[\ref{eq:fitA}]-[\ref{eq:fitB}]) that connects them. Clearly, the observed weak energy dependence of the various Fourier components can be accounted for by simply assuming that the fractional pulse amplitude of each spectral component does not depend on energy. Moreover, the sign change of the fundamental component of the sine Fourier series at energy $\sim 1.7$~keV clearly follows the increase in the flux contribution of the power-law component, which shows substantial pulsations.

\autoref{fig:corner} shows the resulting posteriors for the parameters $f_1 - f_4$ and $\phi_1 - \phi_4$ for $n=1,2$. \autoref{tab:res} contains the best-fit parameters together with their 68\% credible intervals. As expected, the angular arguments $\phi_{in}$ of the surface spectral components do not depend on the harmonic number, i.e., the blue and red contours for $\phi_1$-$\phi_3$ overlap. However, this is not true for the power-law component, which is thought to be of magnetospheric origin. Interestingly, the inferred azimuths of the three spectral components $\phi_1, \phi_2,$ and $\phi_3$ are very close to each other; suggesting that these surface components are nearly co-located.  
\autoref{tab:res} allows us to count the maximum number of independent parameters that can be inferred from the current \nicer data for this source. For each surface spectral component, we can measure the temperature and radius of the emitting region from the DC component, the amplitudes of the fundamental and harmonic Fourier components, and their common phase. As a result, the total number of theoretical model parameters of surface emission that can be used to model \src cannot exceed 15. In the next section, we will discuss the broader implications of this result for model fitting and parameter inference from \nicer pulse profiles.

\section{Conclusions}
\label{sec:concl}

We investigated the relation between the amplitudes of the Fourier components of pulse profiles of \src and the X-ray spectral components using \nicer and \xmm data. We found that the Fourier decomposition of the pulse profiles at different energies supports the detection of two Fourier components that are statistically significant. By combining this information with the X-ray spectral data, we determined that at most 15 independent model parameters can be inferred from the combined timing and spectral data from this source. Equally importantly, we determined that the power-law component of the spectrum, which is presumed to be of magnetospheric origin, shows substantial pulsations with significant harmonic structure.

We generalized our analyses to the three additional sources for which \nicer carried out sustained observing campaigns for pulse profile modeling. We show the results in Appendix~\ref{app:others}. For \psrdz, we can also detect up to two Fourier components, but only at energies $E\lesssim 2$~keV. For this source, the first harmonic is stronger than the fundamental, which is consistent with the high inferred inclination. The spectrum of this source, however, shows evidence for a single blackbody and a power law component, with the blackbody making at most a comparable contribution as the power law in the $0.6-1.3$~keV range, and being subdominant everywhere else. Adopting an alternative model with only blackbody components results in a fit that is statistically worse. More importantly, that model leads to an unphysically low value for the interstellar extinction, forcing the hydrogen column density to become practically zero. This situation is similar to the extremely low {\bf most likely} value of the hydrogen column density inferred for \src by \citet{Choudhury_2024} that is mostly unconstrained and largely inconsistent with independent measurements.

For \psrzs, it is barely possible to detect 1 or 2 Fourier components for $0.3 \leq E \leq 1$~keV, at marginal statistical significance. For this reason, we did not perform a spectral analysis for this source. 

Finally, for \psrtt, up to three Fourier components can be detected for energies up to $\sim 1.3$~keV, beyond which there are no detectable source counts above the background. The X-ray spectrum of this source is dominated by a power law at all photon energies: it has a factor of 5 higher contribution than the blackbody at 0.3 keV and 50\% more contribution at 1~keV. 

For all the {\it spectral} analyses performed here, we note that we do not use \nicer data but only \xmm observations. \cite{salmi2024radius-449} and \cite{vinciguerra2024updated-088} showed that, when simultaneously used, \nicer and \xmm data can also be fit with additional thermal components, which could imply temperature gradients across the hot spots rather than the presence of non-thermal emission. However, neither study provided a direct comparison of the fit quality between the thermal and non-thermal models. 

There are two conclusions that can be reached from this broader analysis. The first relates to the treatment of the power-law component. In previous studies of \nicer targets, the presence of the power-law component has either been neglected or has been incorporated as an unmodulated component of the background 
\citep{riley2019nicer-f21, Choudhury_2024}. In \src, \citet{2016MNRAS.463.2612G} demonstrated conclusively that the power-law component shows significant pulsations. Our analysis here confirms and, in fact, requires the presence of pulsations in the power-law component, even at energies as low as 1~keV. This invalidates the assumption in the previous studies and is likely to change the resulting values for the neutron-star radius for this source, and possibly also for the sources discussed in the Appendix. 

The second conclusion is related to the information content of the pulse profiles and spectra and the number of parameters that can be robustly inferred from theoretical modeling. {\it Each} thermal surface component included in pulse-profile models requires a minimum of 4 parameters, which are the size, temperature, and the two position angles of the hotspot on the stellar surface, assuming the simplest case of a circular, uniform-temperature spot. In addition, the neutron star introduces two baseline parameters, which are its compactness and the observer's inclination with respect to its spin axis, without counting the distance and hydrogen column density to the source or the beaming pattern of radiation from the surface, all of which are assumed to be known or modeled independently. 

On the measurement side, our analysis has shown that each spectral component typically provides 5 constraints: the flux and temperature of each surface component as well as two amplitudes and 1 phase angle from the Fourier analysis of the pulse profiles. This means that, with the present quality of the data, sources that show one thermal spectral component do not provide sufficient information to infer surface emission and neutron star parameters, even when adopting the simplest model for that emission and neglecting all other assumptions and uncertainties. For a source with three spectral components, such as \src, which provides 15 measurable quantities, there is sufficient information to infer both the parameters of three simple hotspots as well as of the neutron star, which require 14 parameters to describe. However, the beaming of radiation, the column density and distance, as well as the presence of the pulsed power-law component can add significant systematic uncertainties and biases that may be difficult to account for. 

One final point to discuss is the advantage of using data from multiple observatories in combined pulse profile and spectral modeling, and the resulting parameter inference. With its precise timing capabilities and larger effective area, \nicer XTI is an excellent instrument for studying energy-dependent pulse profiles of nearby pulsars that show thermal components in their spectra.  However, its non-imaging nature and the relatively low count rates of the target pulsars result in non-negligible and time varying contribution from various background sources. \citet{Choudhury_2024}, who studied the effects of background and used \xmm data for cross-checking, found that the background emission estimated using the 3C50 background model \citep{Remillard2022AJ....163..130R} becomes comparable to source counts at $\sim 1.5$~keV, as well as below 0.3~keV (see also \autoref{tab:charac}). We note that these limits also coincide with the energy range where a non-thermal component becomes easily detectable (in the higher energies) and a larger but cooler blackbody emission becomes dominant (in the low energies, see \autoref{fig:sp_comp}). 

Similar issues are also visible in the energy-dependent pulse profiles shown in Appendix~\autoref{app:others} for the other pulsars, \psrdz, \psrtt, \psrzs. \xmm, in contrast, provides significantly more reliable phase-averaged spectral data, owing to its much lower background and broader spectral sensitivity. Therefore, obtaining the spectral components from the \xmm data and using their fractional contribution to the total emission, under the assumption that these components do not show significant changes within the \nicer observations used here, is likely to lead to smaller uncertainties than using \nicer data alone.

\section*{Acknowledgements}
T.G. greatly appreciates the hospitality at the School of Physics at the Georgia Institute of Technology, where this work was completed. We thank the anonymous referee for useful comments.

\section*{Data Availability}
All the data used in this paper are publicly available
through the NASA/HEASARC archives. The data from \cite{Choudhury_2024} is available through Zenodo services.

\bibliography{references}
\bibliographystyle{aasjournal}

\appendix
\restartappendixnumbering
\section{Additional Details of the Fourier Analysis}
\label{app:det_fig}

\begin{figure}
    \centering
    \includegraphics[scale=0.5]{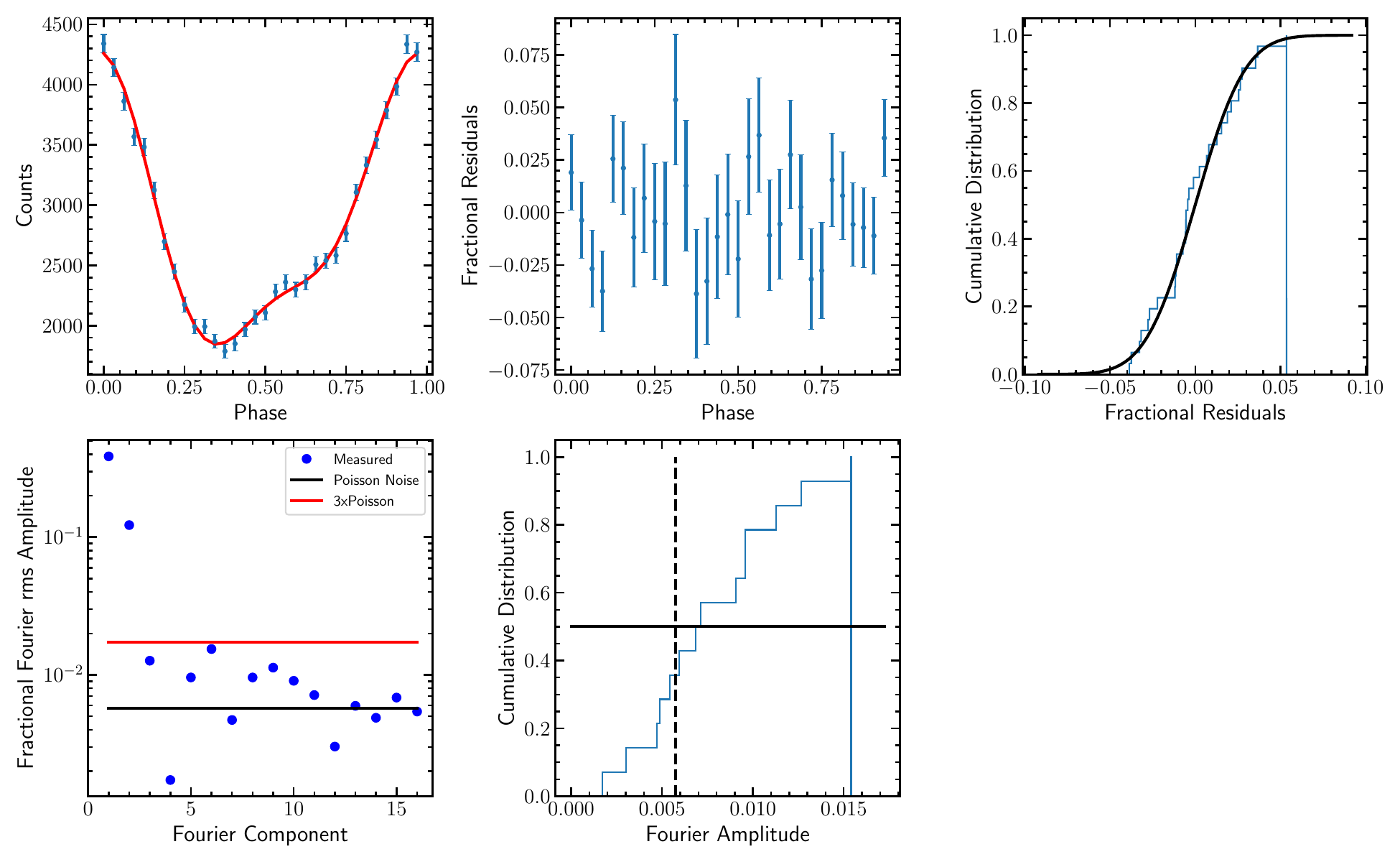}
    \caption{The 0.8--0.9~keV pulse profile of \src obtained with \nicer, as a representative example. From upper left to lower right, the panels show: the pulse profile (blue points) and the model neglecting all Fourier components with amplitudes below 3 times the Poisson noise level (red curve); the residuals from neglecting these Fourier components, showing no systematic dependence on pulse phase; the cumulative distribution of these residuals comparing them also to the expected normal distribution (black curve); the fractional rms amplitudes of the Fourier components together with the expected level of the Poisson noise (black) and 3 times that level (red); and the cumulative distribution of the Fourier amplitudes, demonstrating that the median of this distribution (horizontal black line) closely matches the expected value.}
    \label{fig:detail_profile}
\end{figure}

In this appendix, we provide additional details and tests of our pulse profile analysis. In \autoref{fig:detail_profile}, we show the results of our detailed analysis of the pulse profiles from \src, using the 0.8-0.9~keV range as a representative example. In addition to the pulse profile and the Fourier components, we also show the residuals from neglecting the Fourier components with amplitudes below 3 times the Poisson level, the cumulative distribution of these residuals, as well as the distribution of the Fourier amplitudes of the components that are within 3 times the Poisson noise level. Neglecting these Fourier components results in residuals that show no systematic dependence on pulse phase and with a distribution that closely follows the expected normal distribution. Moreover, the median of the cumulative distribution of Fourier amplitudes closely matches the expected value calculated using eq.~(A6) of \cite{psaltis2014}. All these suggest that there is no evidence of any systematics in our analysis and that our thresholds accurately describe the noise level of the measurements.

\begin{figure}
    \centering
    \includegraphics[width=0.3\linewidth]{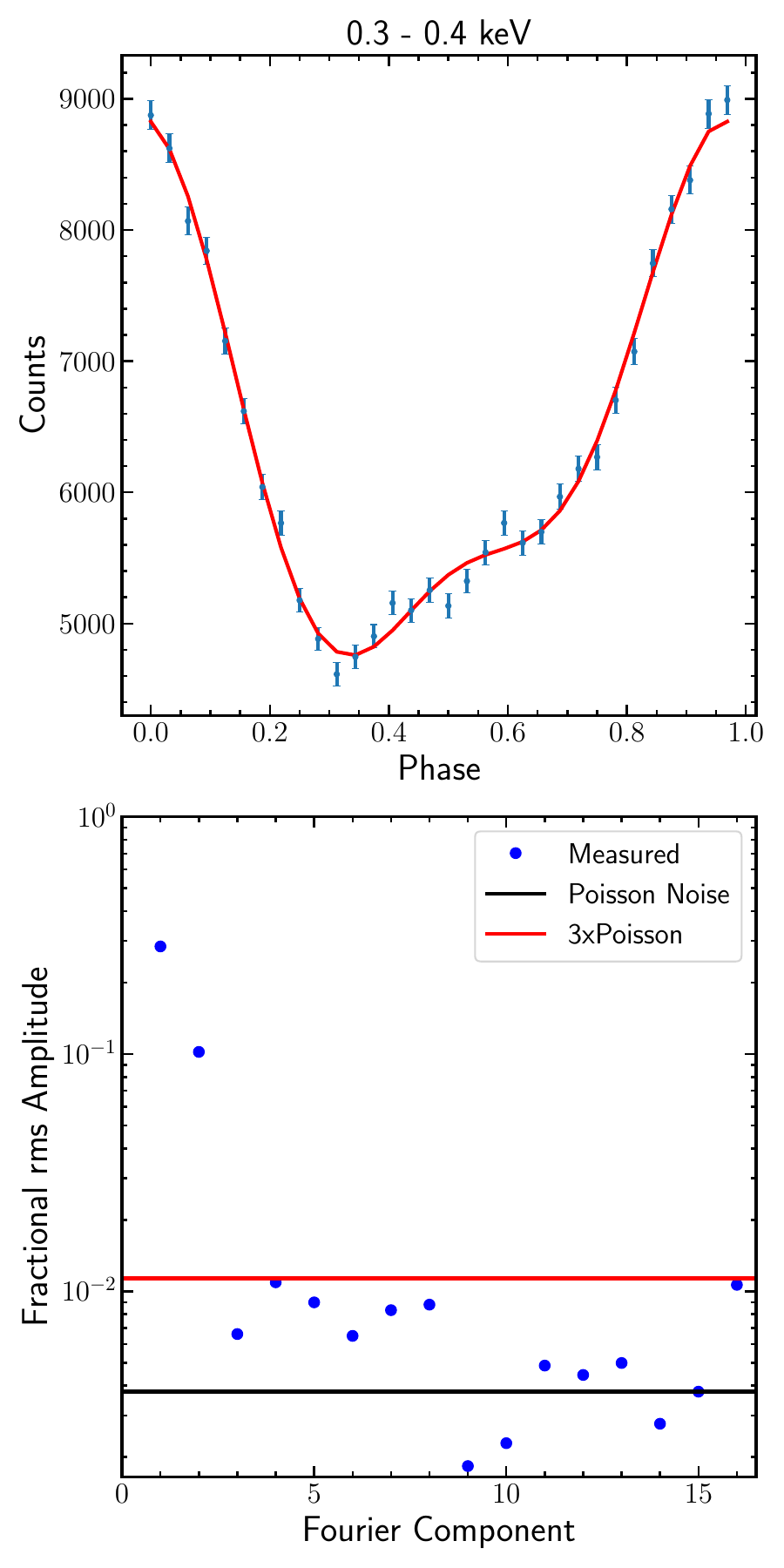}
    \includegraphics[width=0.3\linewidth]{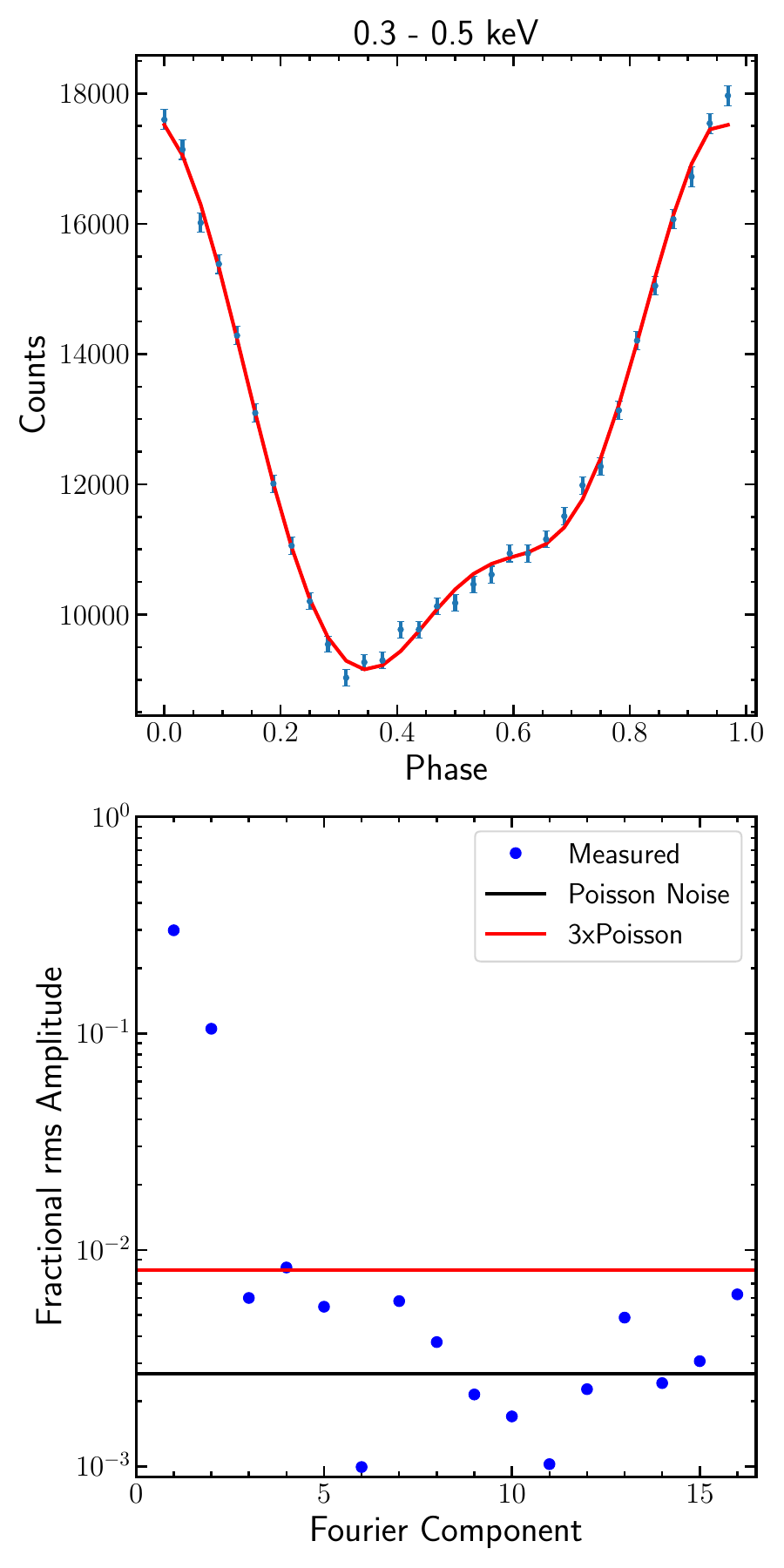}
    \includegraphics[width=0.3\linewidth]{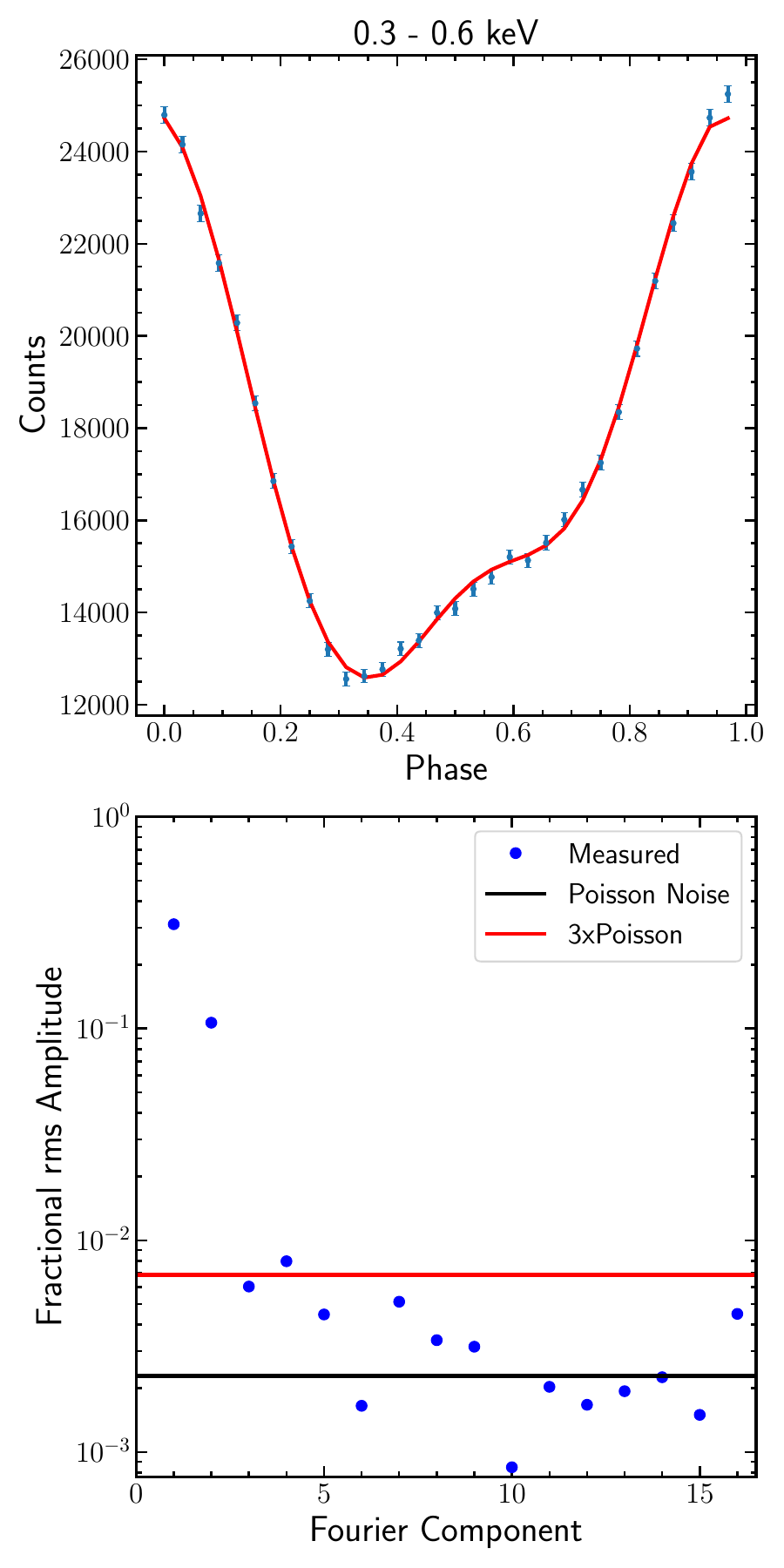}

    \caption{ Same as \autoref{fig:pp1} but showing the effect of increasing the width of the energy range over which we perform the Fourier decomposition.}
    \label{fig:energ_intervals}
\end{figure}

We also test the effects of the selection of energy intervals for pulse profile analysis. Specifically, in Fig.~\ref{fig:energ_intervals}, we show the results of the analysis for different widths of the energy range, i.e., the effect of broadening the energy range we take into account to calculate the pulse profiles and Fourier components. For illustration, we extend the range from 0.3--0.4~keV to 0.3--0.6~keV. There is negligible effect on the number of components that are detectable and on their inferred amplitudes, for reasonable extents of the energy range, before they become wide enough to encompass different spectral components.

\section{Use of Neutron Star Atmosphere Models}
\label{app:nsa_app}

In several previous studies, pulse profile modeling has incorporated neutron star atmosphere models in addition to simple blackbody approximations \citep[see, e.g.,][]{Choudhury_2024}. In this appendix, we perform a similar assessment using the \emph{nsatmos} model \cite{2006ApJ...644.1090H,2004ApJ...615..402M} in XSPEC. Because neutron star atmospheres account for the energy-dependent opacity of fully or partially ionized H, He, or heavier elements, they generally produce broader spectra than pure blackbodies \citep[see, e.g.,][]{2002nsps.conf..263Z,2013RPPh...76a6901O}. Consequently, thermal components modeled with atmospheres yield relatively more emission at higher photon energies. Here, we evaluate how the use of \emph{nsatmos} model affects the results presented in the main text.

For the X-ray spectral data, we adopt the best-fit parameters from the joint ROSAT, \xmm, and \nustar\ analysis presented by \citet{2016MNRAS.463.2612G}, allowing only the model normalizations to vary. The fit included one blackbody component for the surface emission (kT$_1=27_{-4}^{+5}$~eV), two fully ionized hydrogen atmosphere components for the hot spots (kT$_2 = 51\pm$6~eV, kT$_3=147_{-8}^{+11}$~eV), and a non-thermal power-law component ($\Gamma = 1.50 \pm 0.25$, normalization 1.4$_{-0.5}^{+0.7}$). The corresponding hydrogen column density from this fit is N$_{\rm{H}}=3.3_{-1.3}^{+1.6}\times10^{20}~\rm{cm}^{-2}$. 

\begin{figure}
    \centering
    \includegraphics[width=1.0\linewidth]{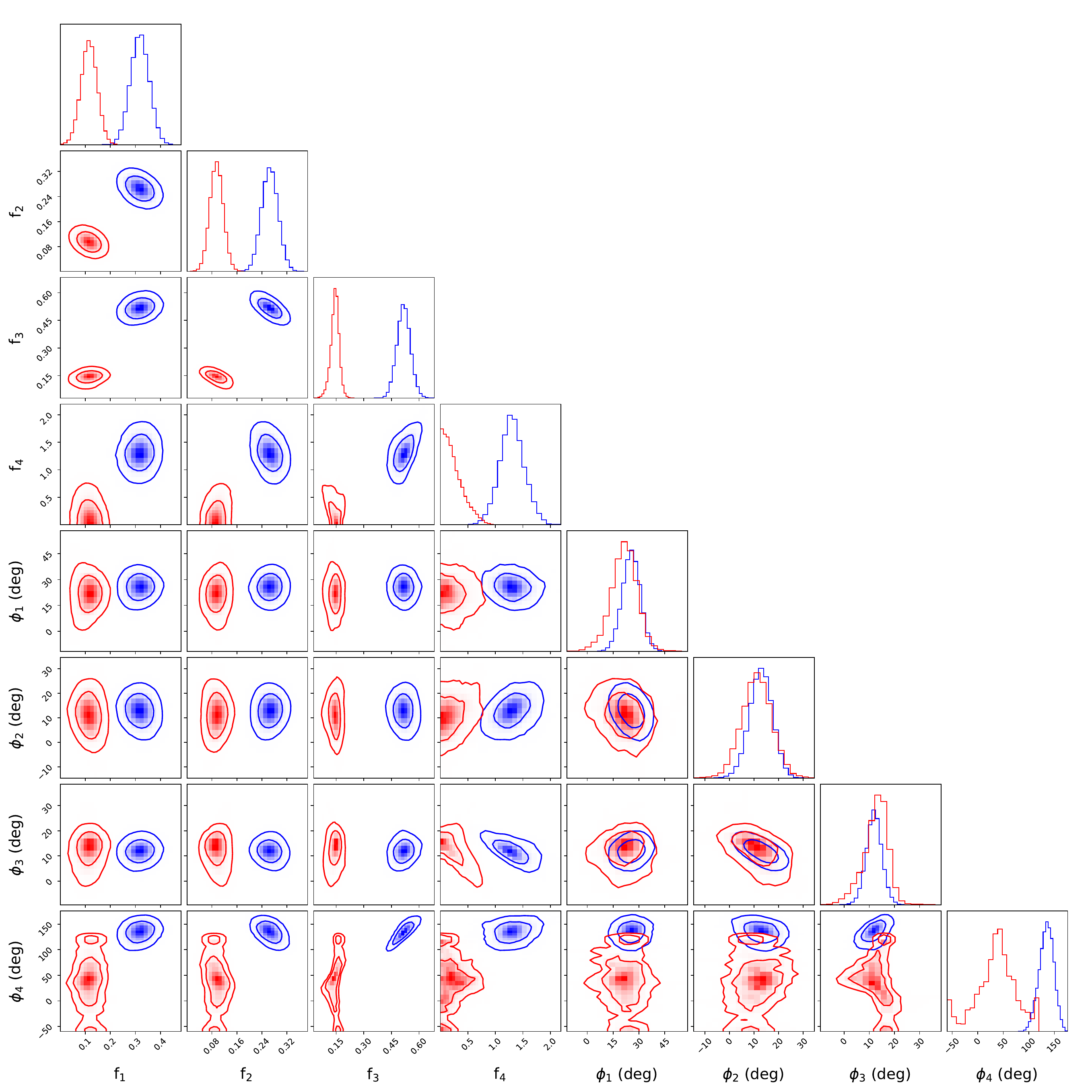}
    \caption{Same as \autoref{fig:corner} but assuming that the two thermal components are neutron star atmosphere models and not blackbody functions.}
    \label{fig:corner_nsa}
\end{figure}

Using the same pulse profile data as in Section~\ref{sec:pulse_profiles} and the fractional fluxes from these models, we obtain the results shown in \autoref{tab:nsa_res} and \autoref{fig:corner_nsa}. These results indicate that the use of the atmosphere model slightly alters the energy in which the non-thermal component produces comparable fluxes.

\begin{table*}[ht!]
\centering
\caption{Spectral components assuming \emph{nsatmos} models and their corresponding contributions to Fourier amplitudes.}
\begin{tabular}{cccccc}
\hline
&&\multicolumn{2}{c}{Fundamental} & \multicolumn{2}{c}{Harmonic}\\
Spectral Component & Parameters$^{1}$ & Amplitude& Phase & Amplitude & Phase \\
 & kT (eV), R (km)& $f_{i1}$ & $\phi_{i1}$ (deg)& $f_{i2}$ & $\phi_{i2}$ (deg)\\
\hline
Blackbody 1 & 27, 51  & 0.32$\pm$0.04& 25.3$\pm$5.3 & 0.12$\pm$0.03&20.4$\pm$14.5\\
\emph{nsatmos} 2  & 51, 4.5 &0.26$\pm$0.03  &12.9$\pm$4.6& 0.096$\pm$0.02&13.6$\pm$12.1\\
\emph{nsatmos} 3 & 147, 0.4 & 0.52$\pm$0.04 &12.1$\pm$3.00 &0.14$\pm$0.02 &8.9$\pm$9.2\\
\hline
Power-law & $\Gamma$=1.5, 1.4$\times10^{-5}$ & 1.29$\pm$0.22&137.1$\pm$14.8 & 0.30$\pm$0.19 &49.25$\pm$72.9 \\
\hline
\end{tabular}
\label{tab:nsa_res}\\
\footnotesize{$^{1}$ Values are adopted from \citet{2016MNRAS.463.2612G}.}
\end{table*}

\section{Other Sources}
\label{app:others}
He we present our analysis of three additional \nicer targets, namely \psrdz, \psrzs, and \psrtt. 

\subsection{\psrdz}

\begin{figure}
    \centering
    \includegraphics[scale=0.35]{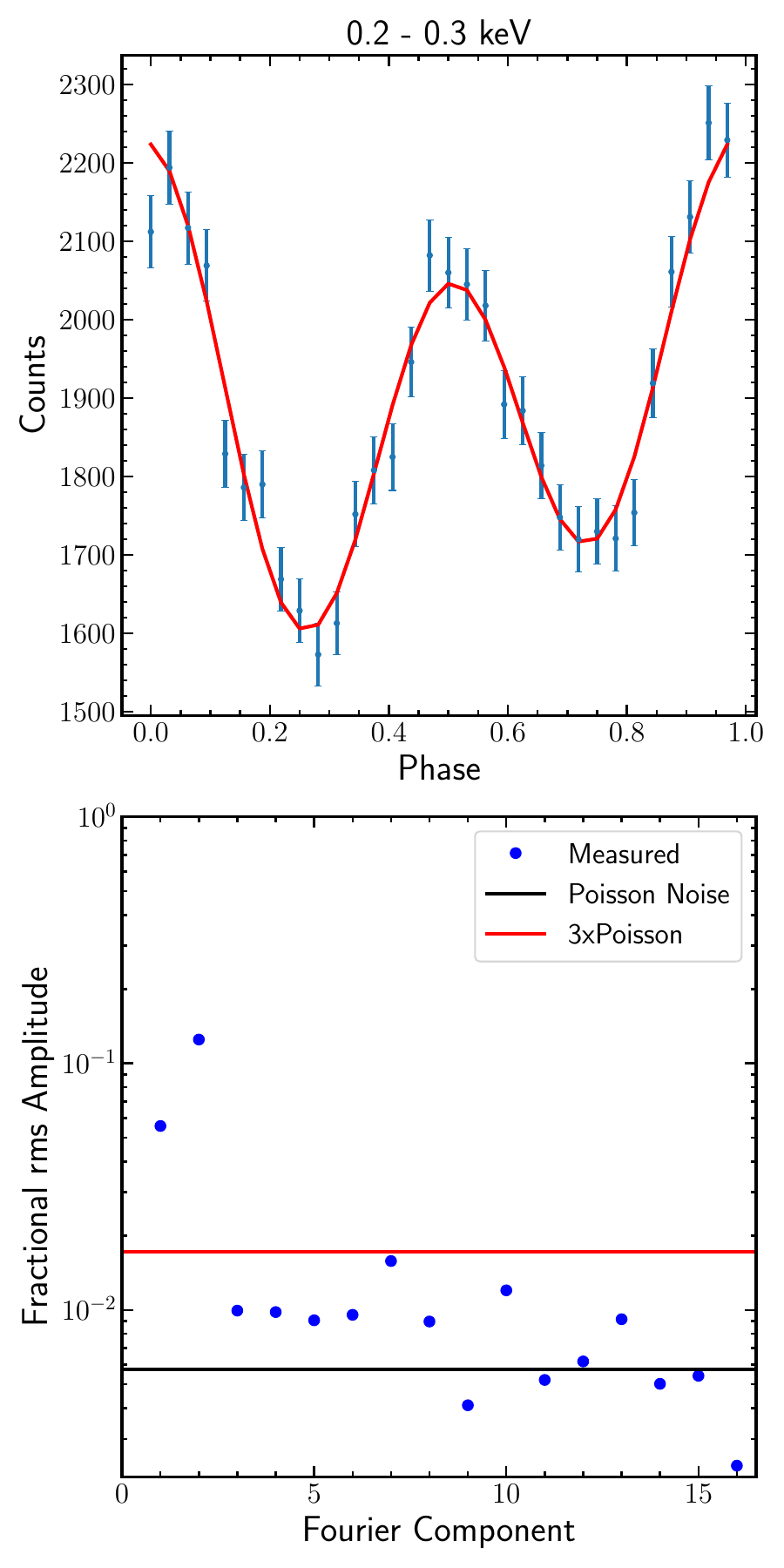}
        \includegraphics[scale=0.35]{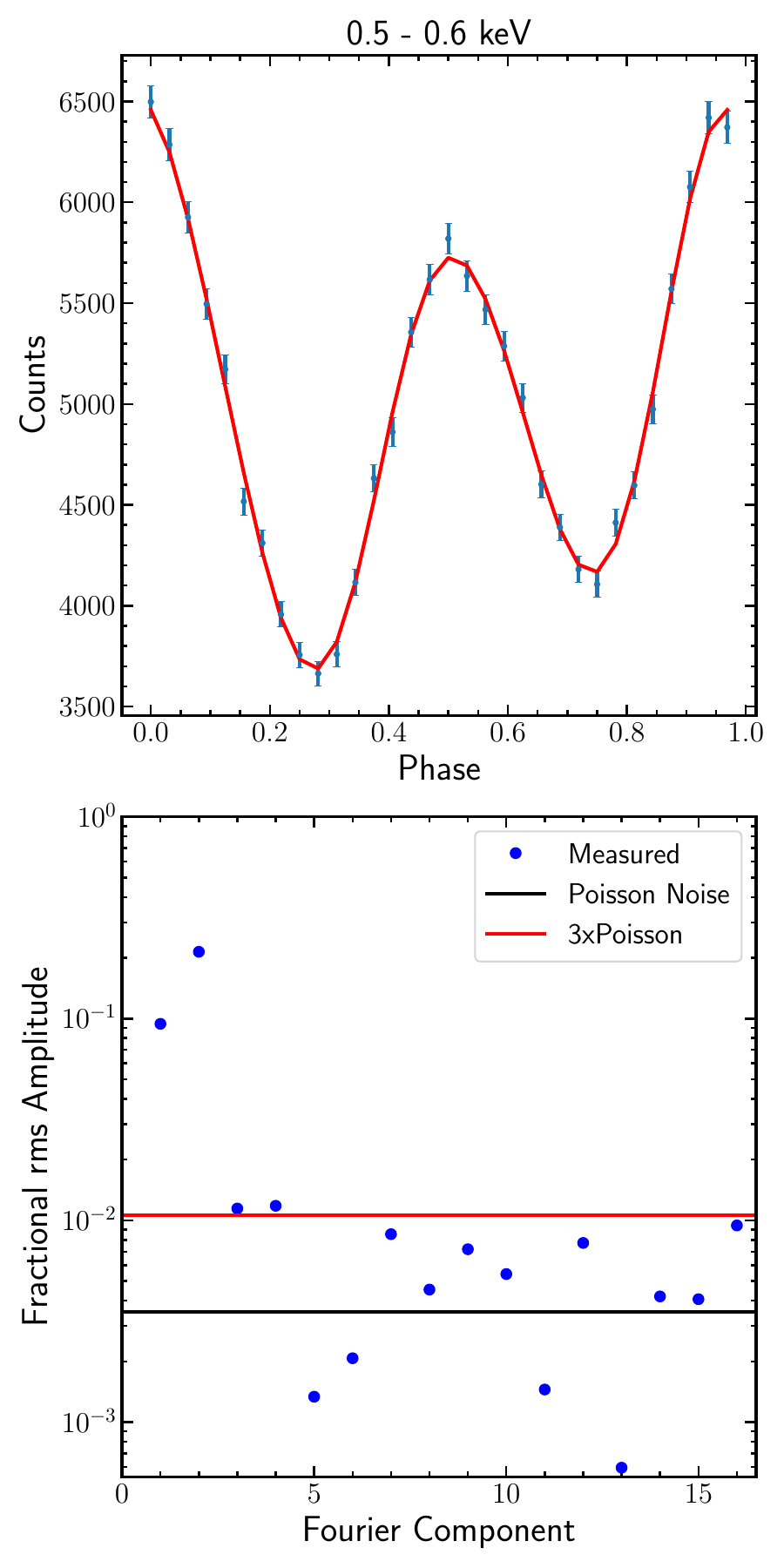}
            \includegraphics[scale=0.35]{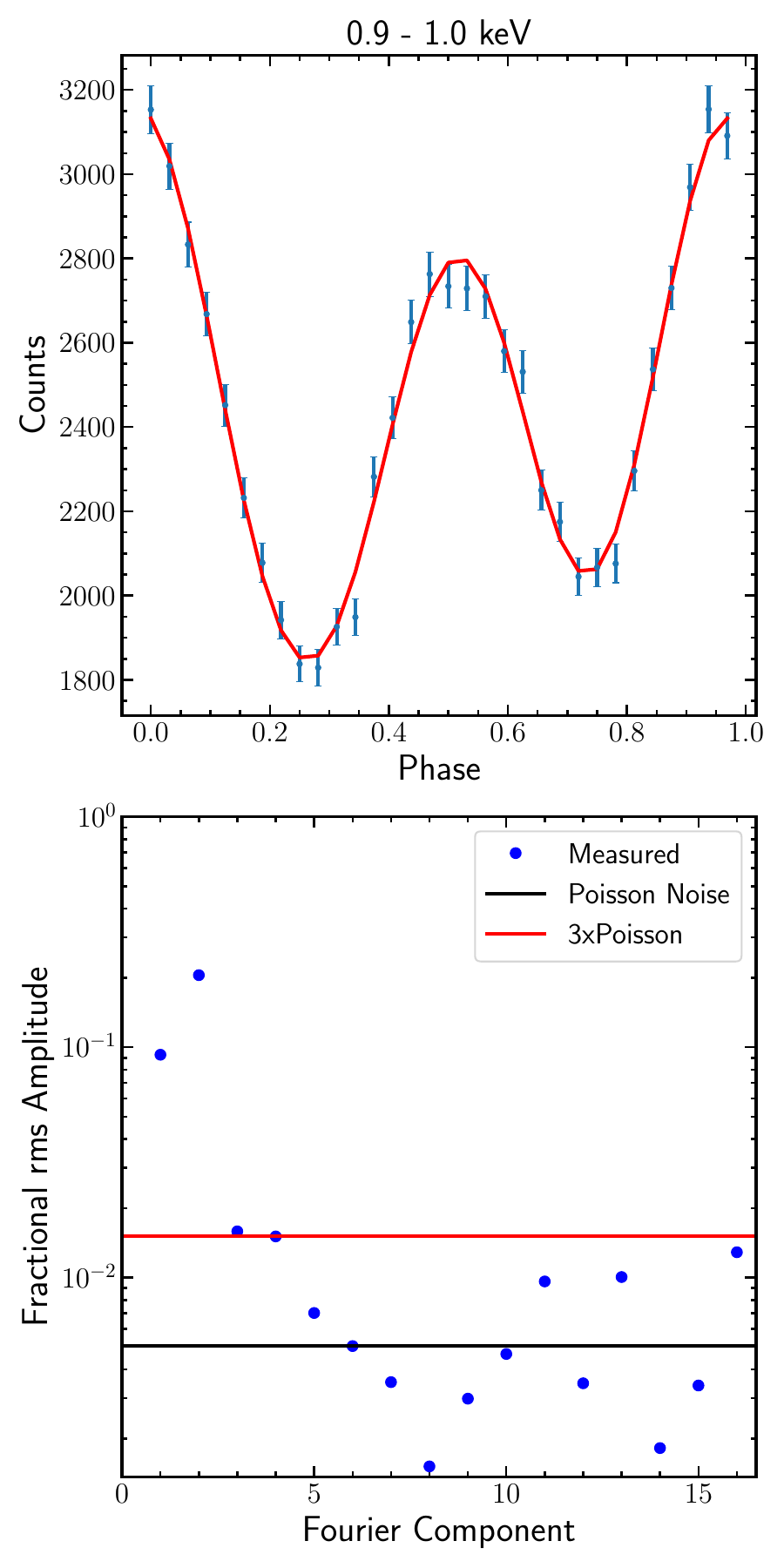}
    \includegraphics[scale=0.35]{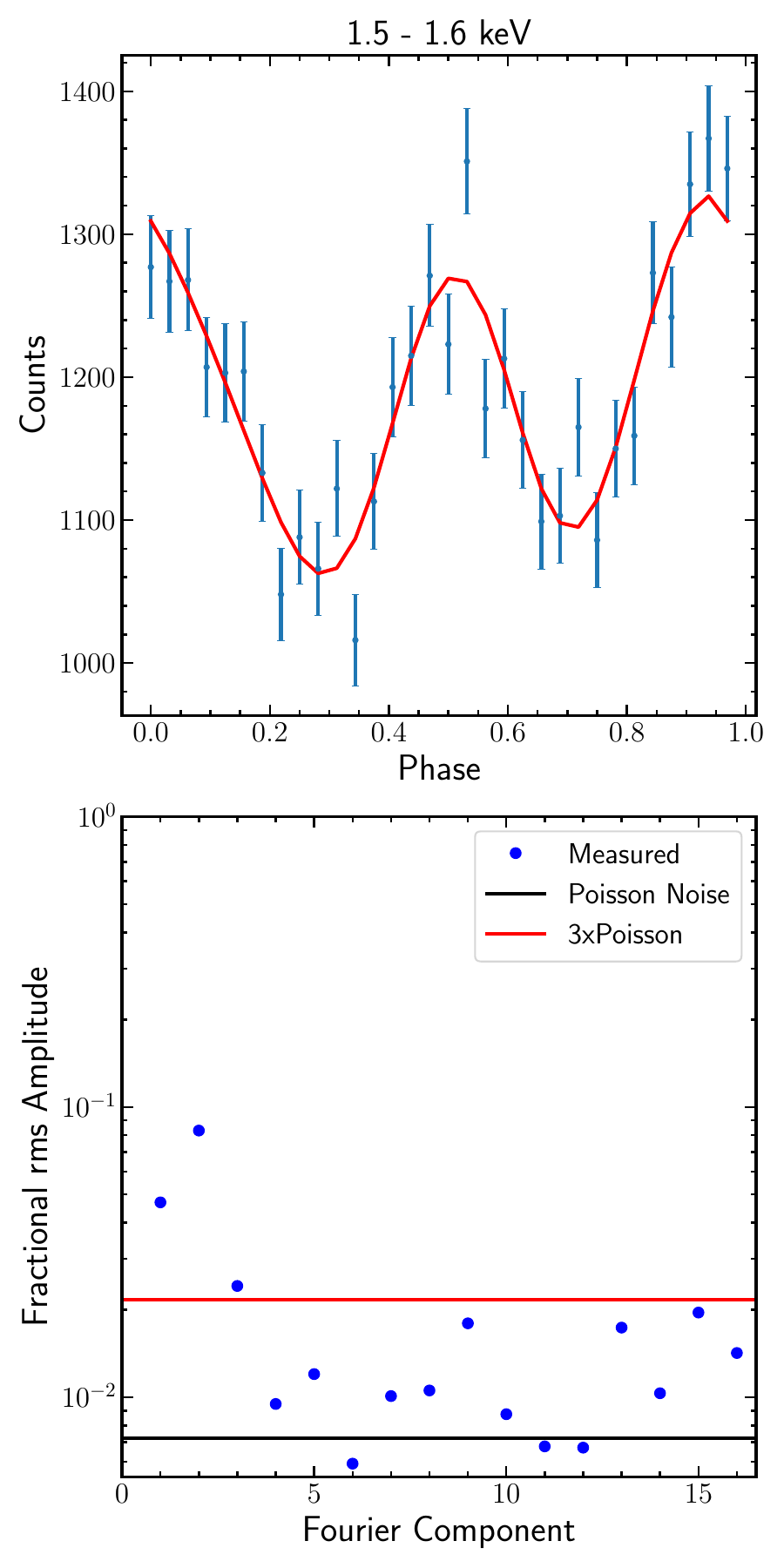}
    \includegraphics[scale=0.35]{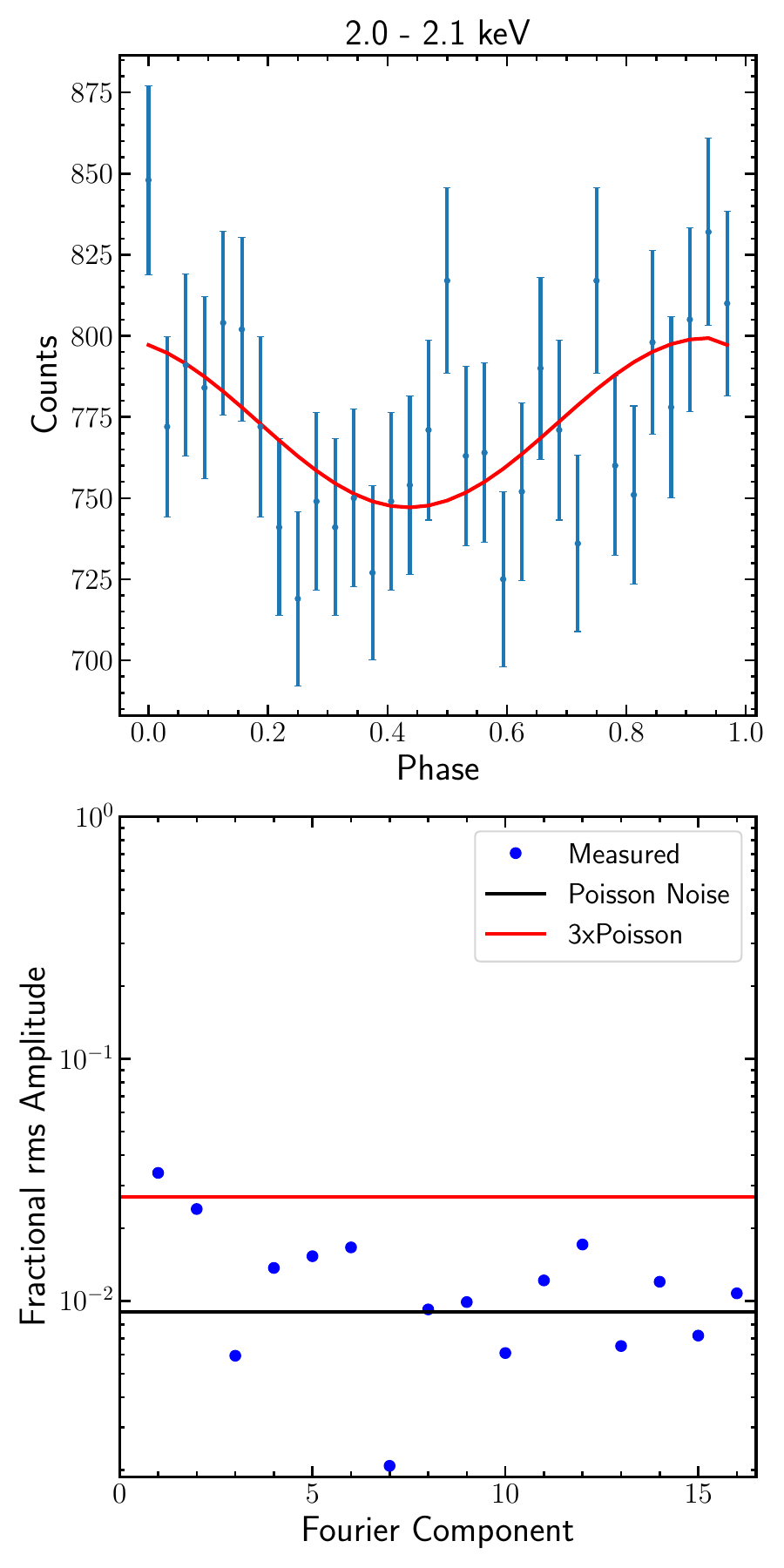}
        \includegraphics[scale=0.35]{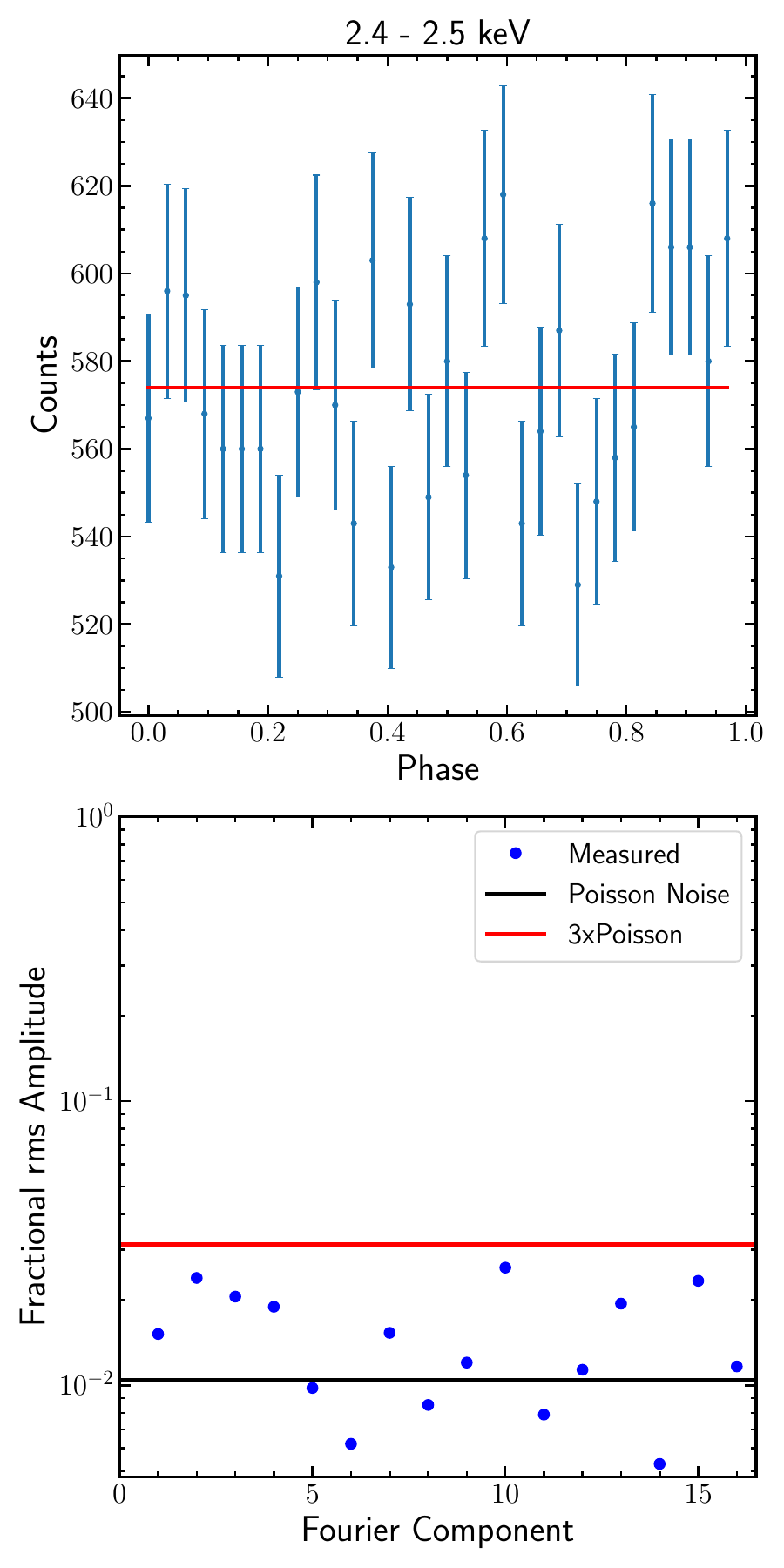}
    \caption{Same as \autoref{fig:pp1} but for \psrdz. Pulses are only detectable below 2.0~keV.}
    \label{fig:psr0030pp}
\end{figure}

In \autoref{fig:psr0030pp}, we show examples of pulse profiles obtained with \nicer at several energy bands, similar to the ones shown in \autoref{fig:pp1} for \src. Note that we did not perform a background subtraction as no background file for \nicer data is present in the Zenodo archives for this source. It is evident that pulsations are detectable only up to 2~keV.

\begin{figure}
    \centering
    \includegraphics[scale=0.4]{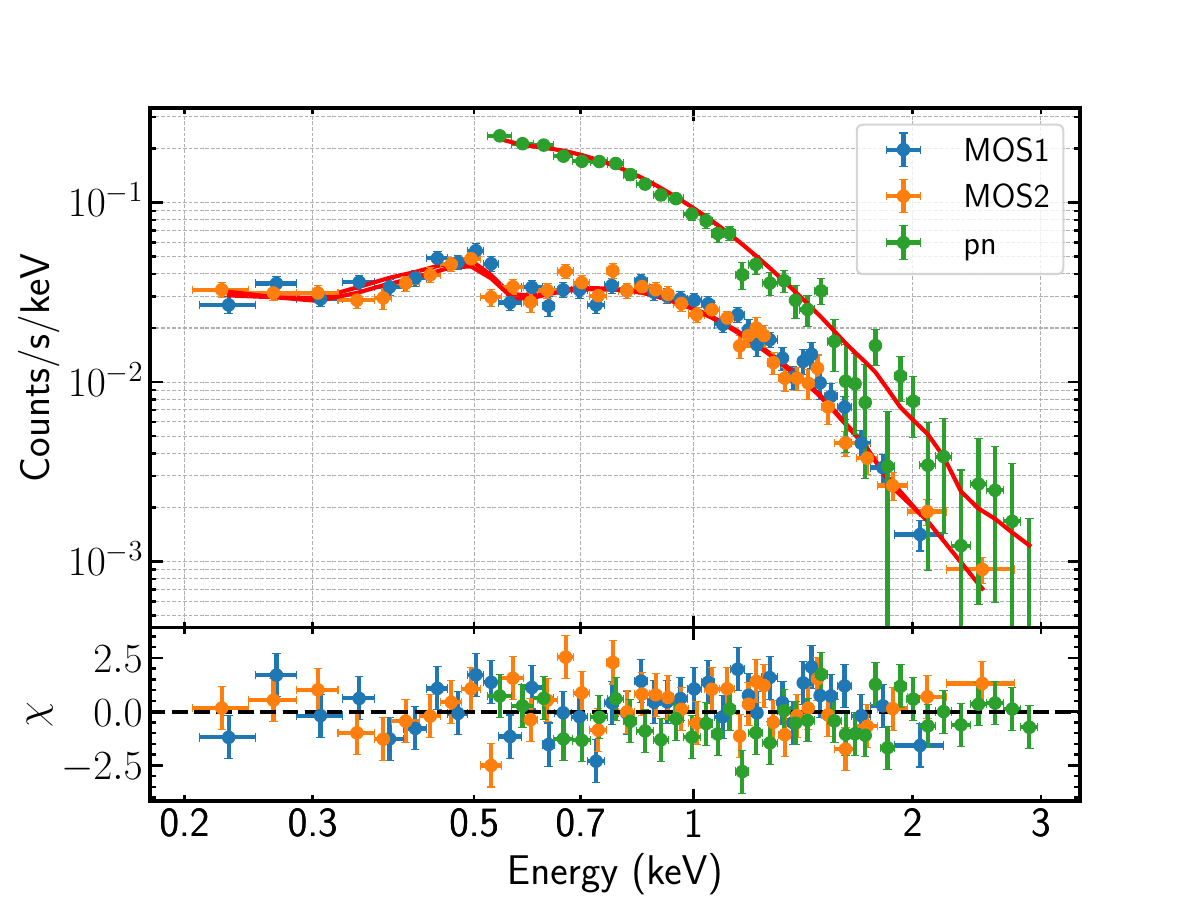}
        \includegraphics[scale=0.4]{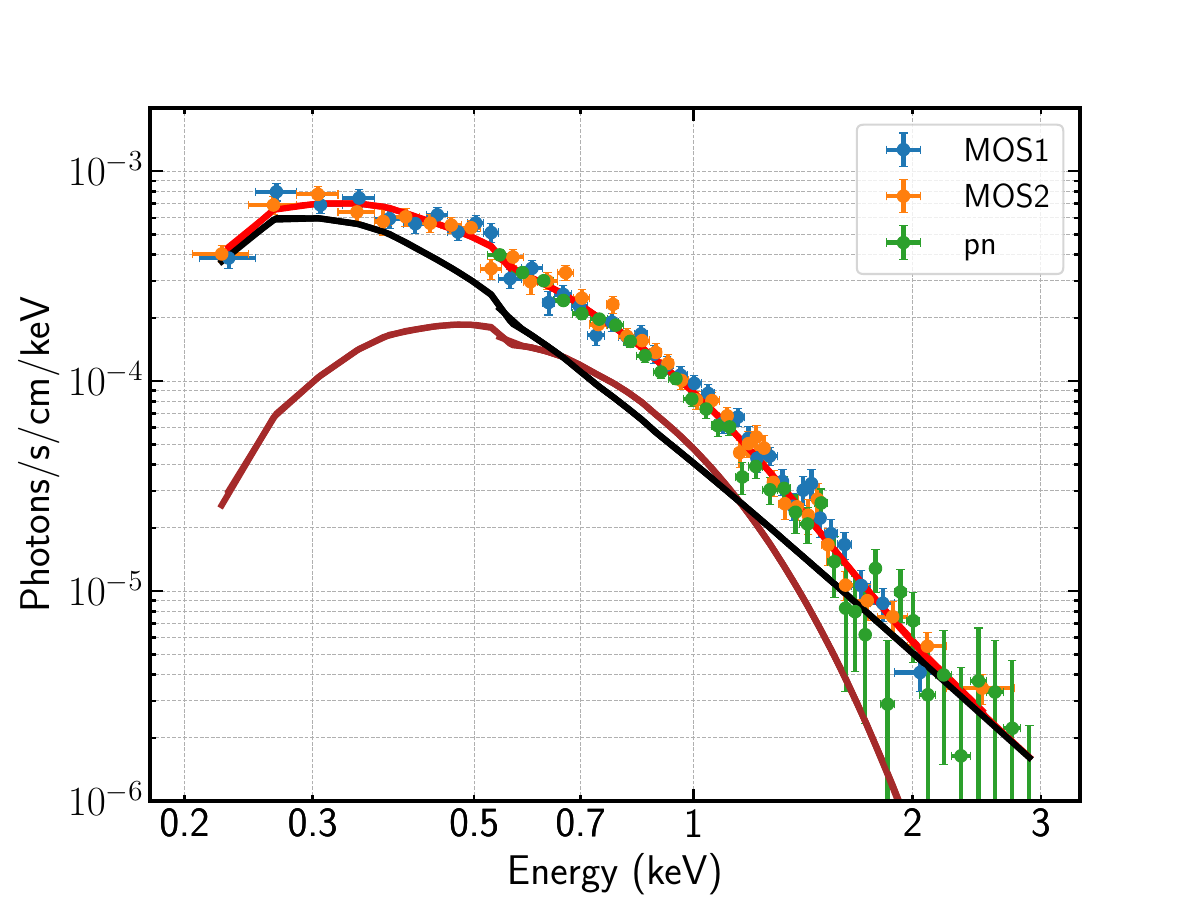}
    \caption{X-ray spectra of \psrdz obtained with \xmm. The left panel shows the data and the best fit model (upper panel) together with residuals (lower panel). The right panel shows the model components, which include an absorbed blackbody (red) and a power-law (black) component.}
    \label{fig:psr0030sp}
\end{figure}

We also show in \autoref{fig:psr0030sp} the spectra of \psrdz obtained with \xmm. The spectra show evidence for only one thermal component and a significant power-law component that dominates over the thermal one at $E\lesssim 0.6$~keV and $E \gtrsim 1.3$~keV (with a $\chi^2$ of 130.13 for 111 degrees of freedom). The X-ray spectra can also be modeled with a combination of two blackbodies, but this fit results in a worse $\chi^2$ of 151.99 for the same degrees of freedom. However, in the latter  case, the hydrogen column density attains an unphysically small value that is consistent with zero.

\subsection{\psrzs}

\begin{figure}
    \centering
    \includegraphics[scale=0.35]{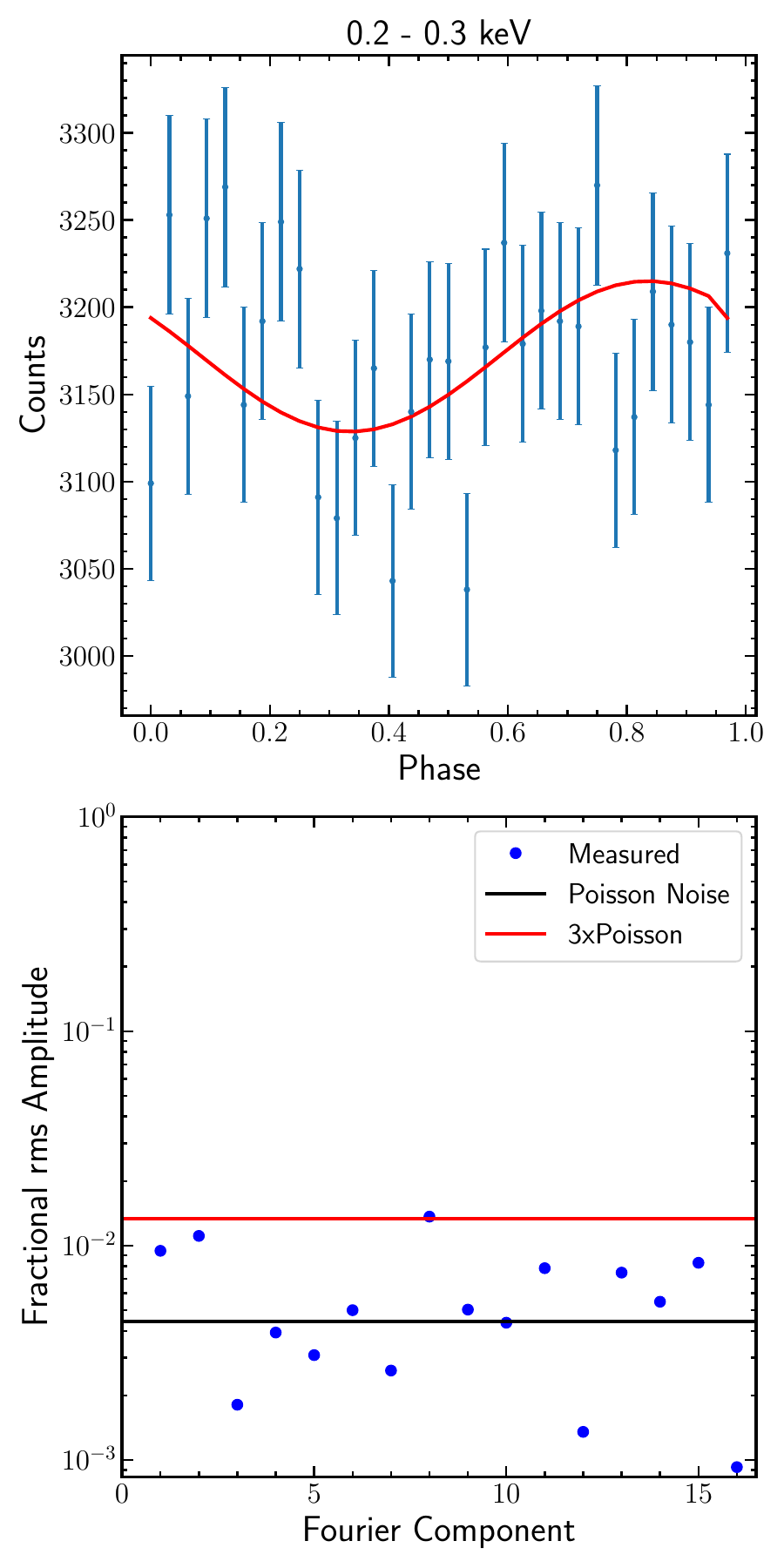}
        \includegraphics[scale=0.35]{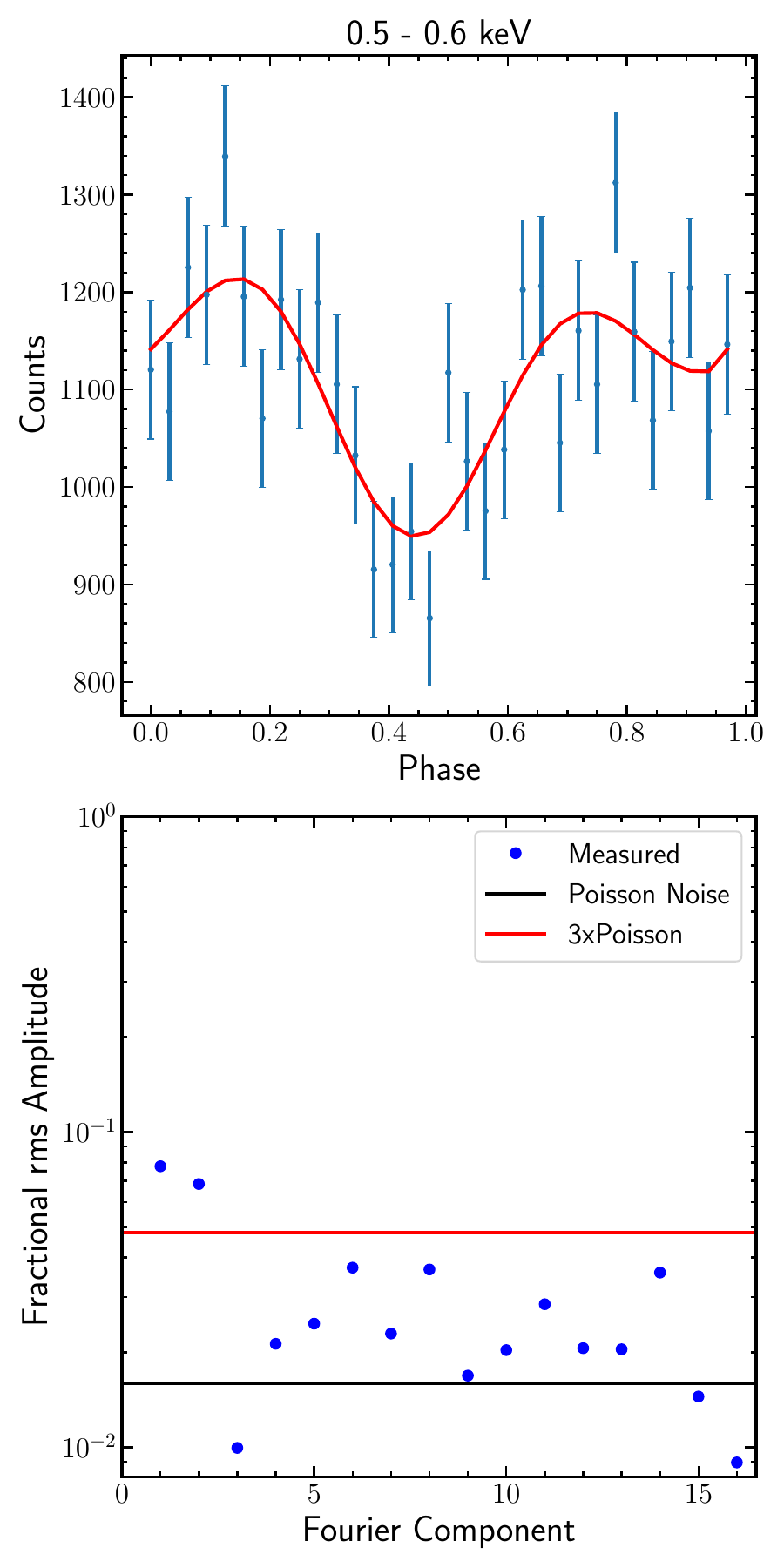}
            \includegraphics[scale=0.35]{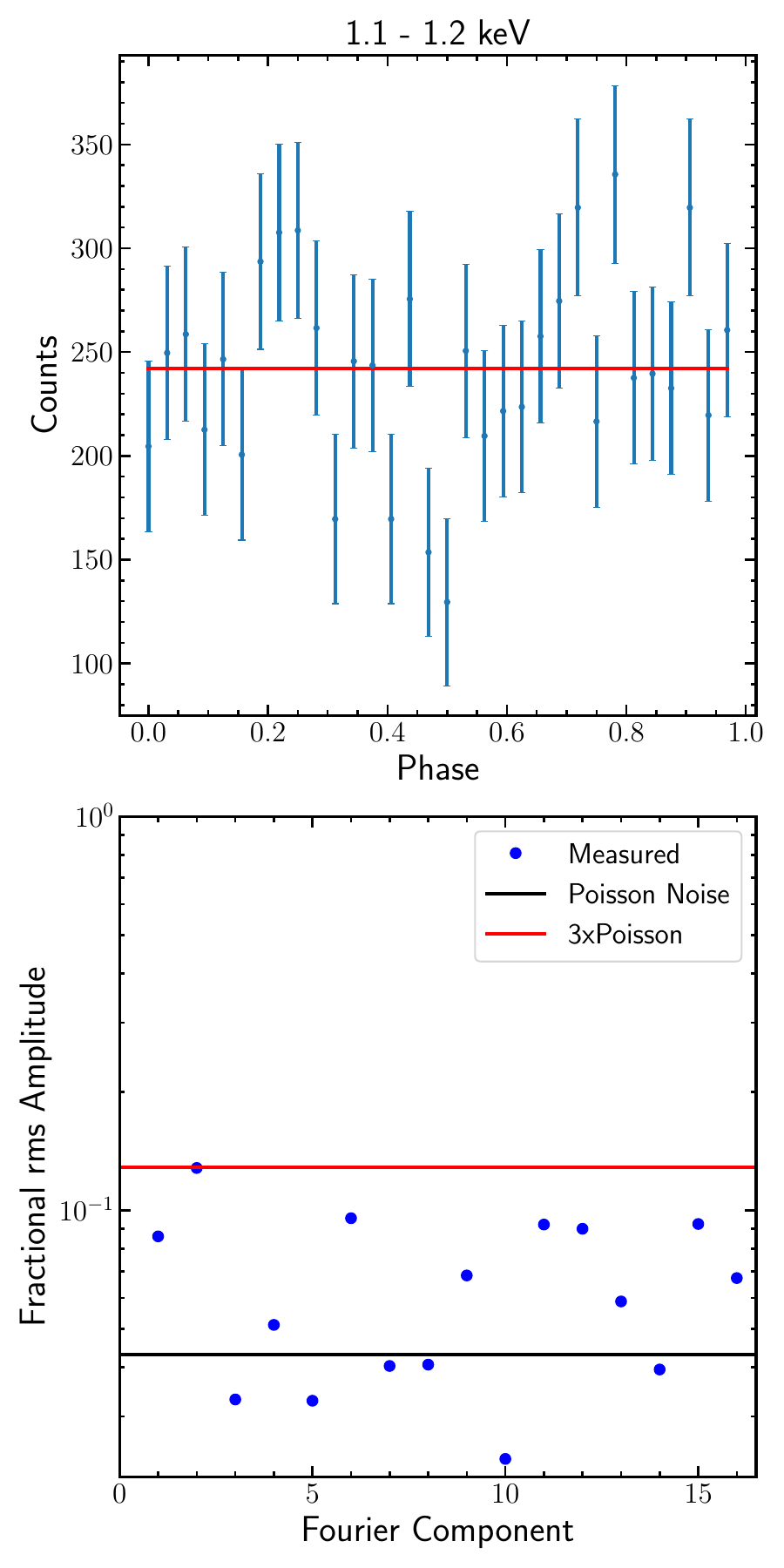}
    \includegraphics[scale=0.35]{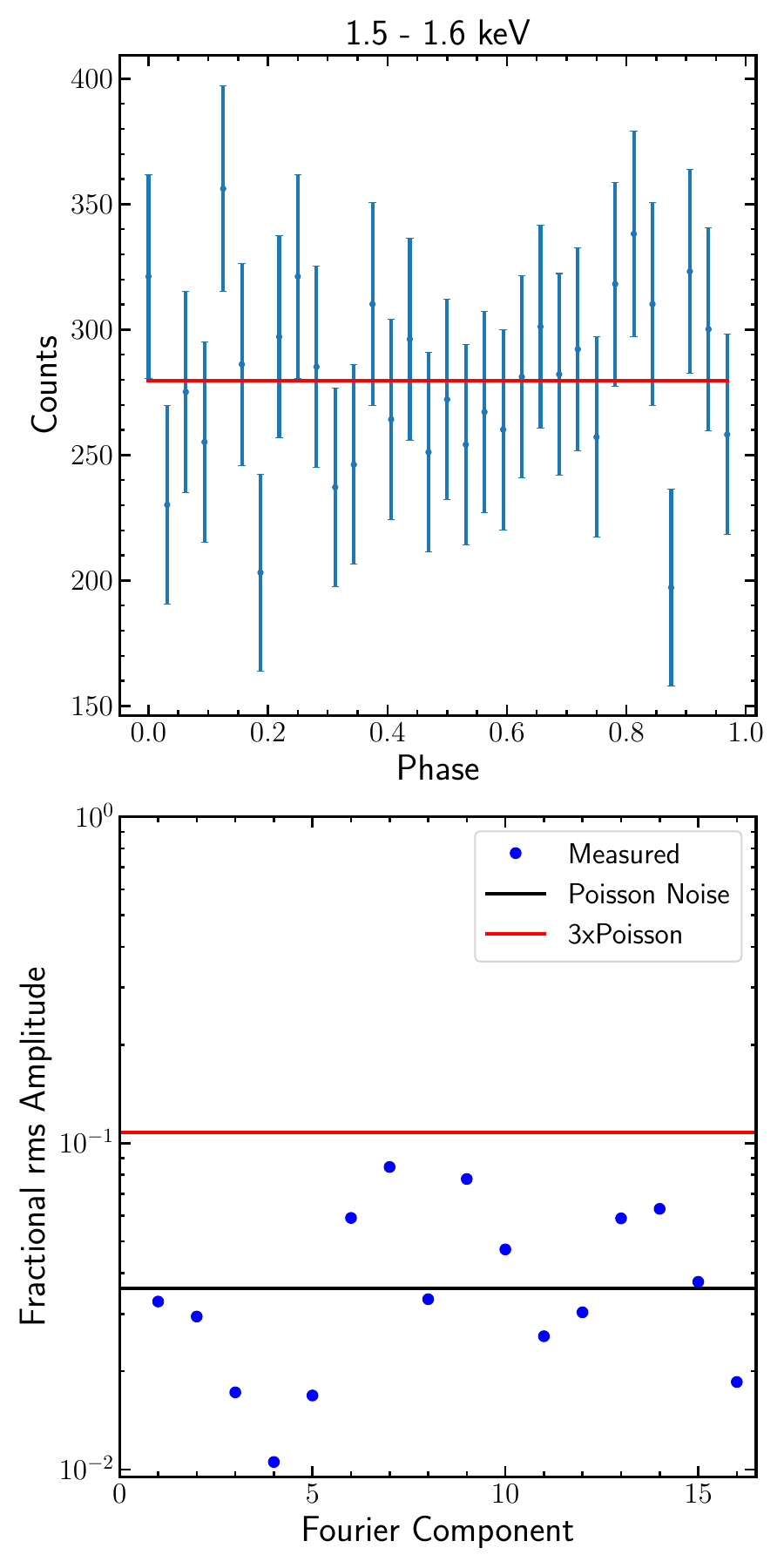}
    \includegraphics[scale=0.35]{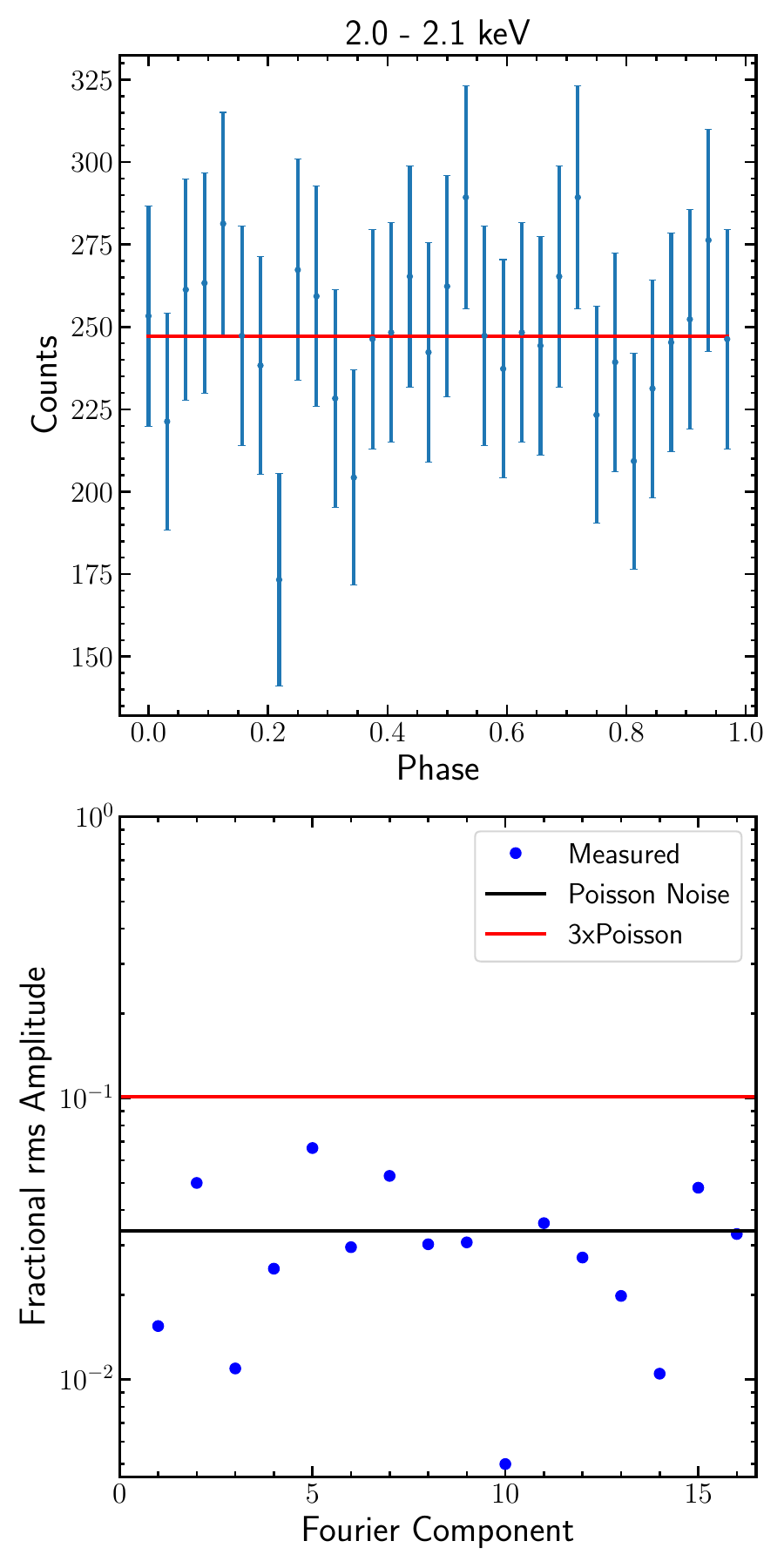}
        \includegraphics[scale=0.35]{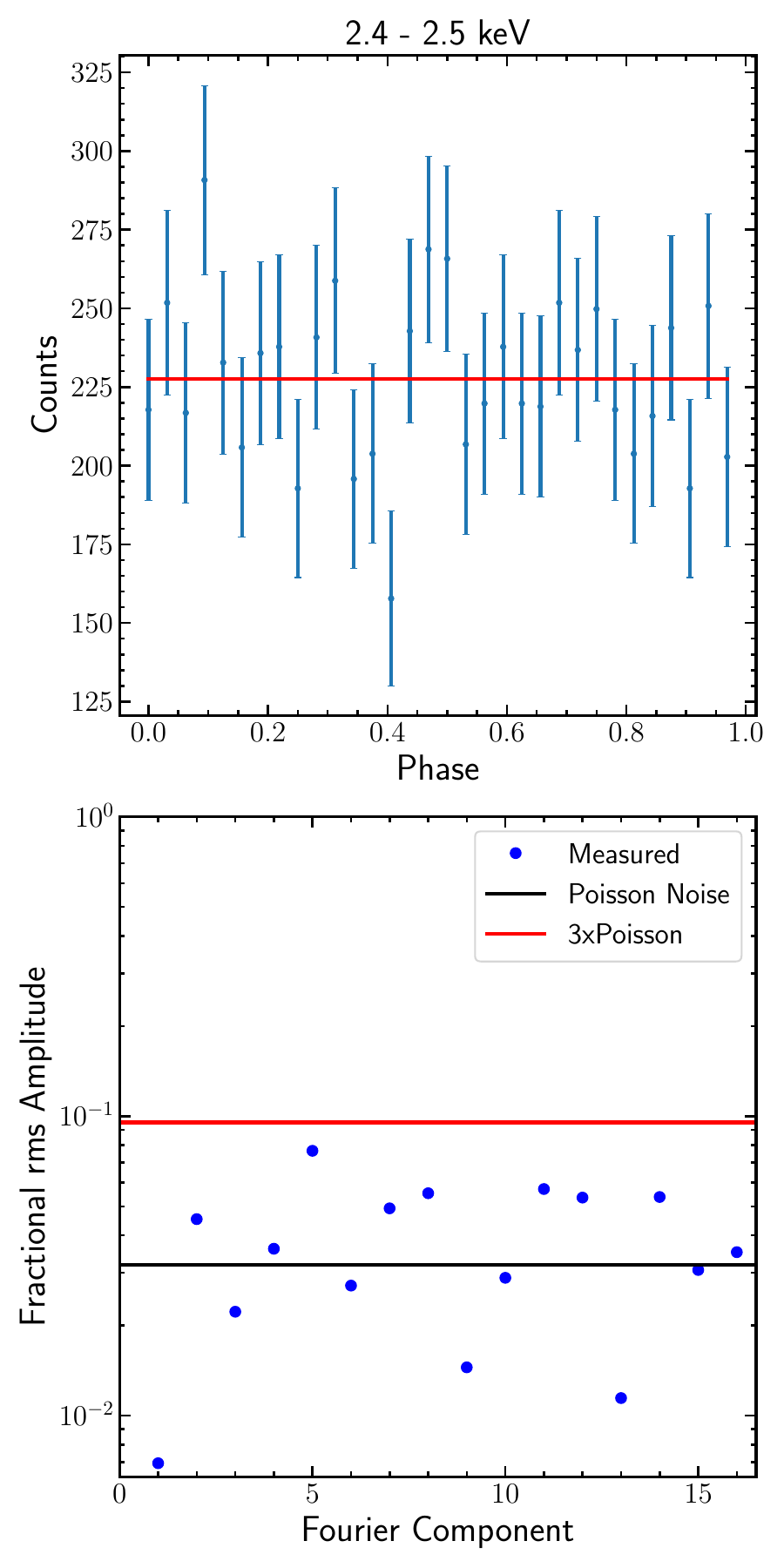}
    \caption{Same as \autoref{fig:pp1} but for \psrzs. For this pulsar, pulsations are only detected between 0.4-1~keV.}
    \label{fig:psr0740pp}
\end{figure}

In \autoref{fig:psr0740pp}, we show examples of pulse profiles obtained with \nicer at several energy bands for \psrzs. The pulsations are (marginally) detectable between 0.3 and 1~keV, with at most 2 harmonics. This source has been detected by \xmm \citep[see e.g.,][]{2021ApJ...918L..28M} but only with a very limited number of counts (in total 220 counts from pn, including MOS1 and MOS2). We, therefore, opt to not include an X-ray spectral analysis here. X-ray spectra can be seen in Figure 8 of \cite{2021ApJ...918L..28M}.

\subsection{\psrtt}

\begin{figure}
    \centering
    \includegraphics[scale=0.35]{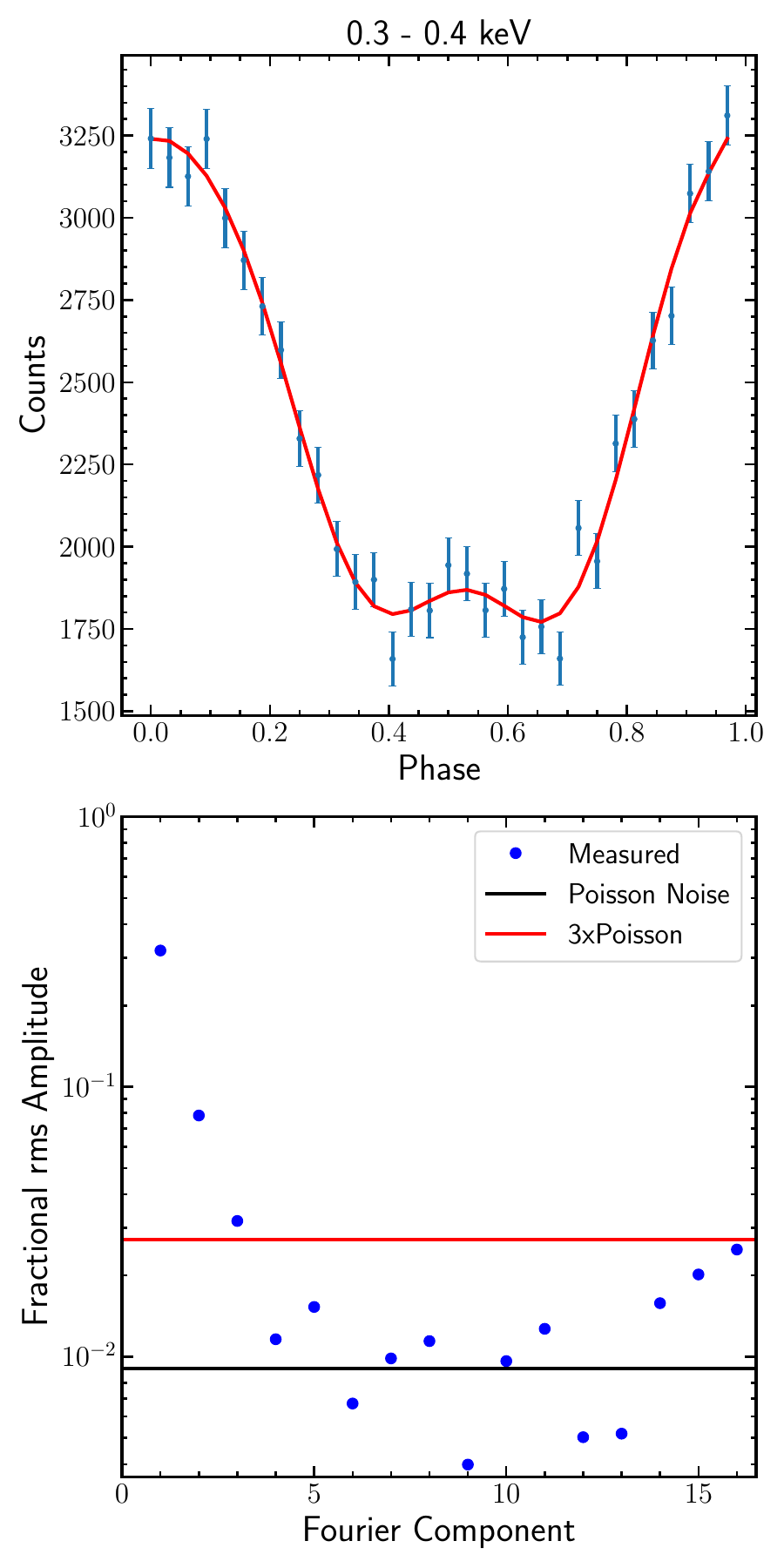}
        \includegraphics[scale=0.35]{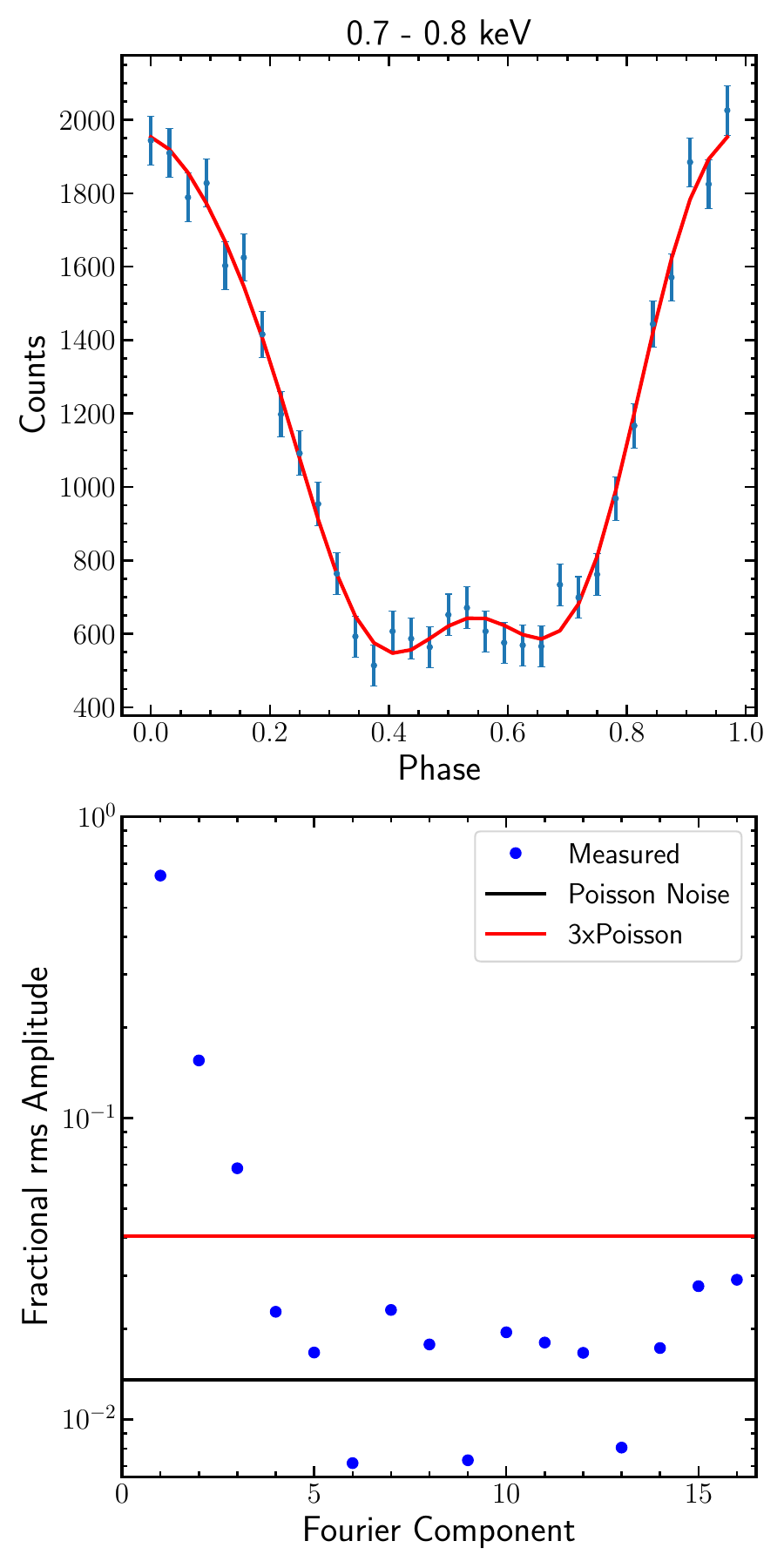}
            \includegraphics[scale=0.35]{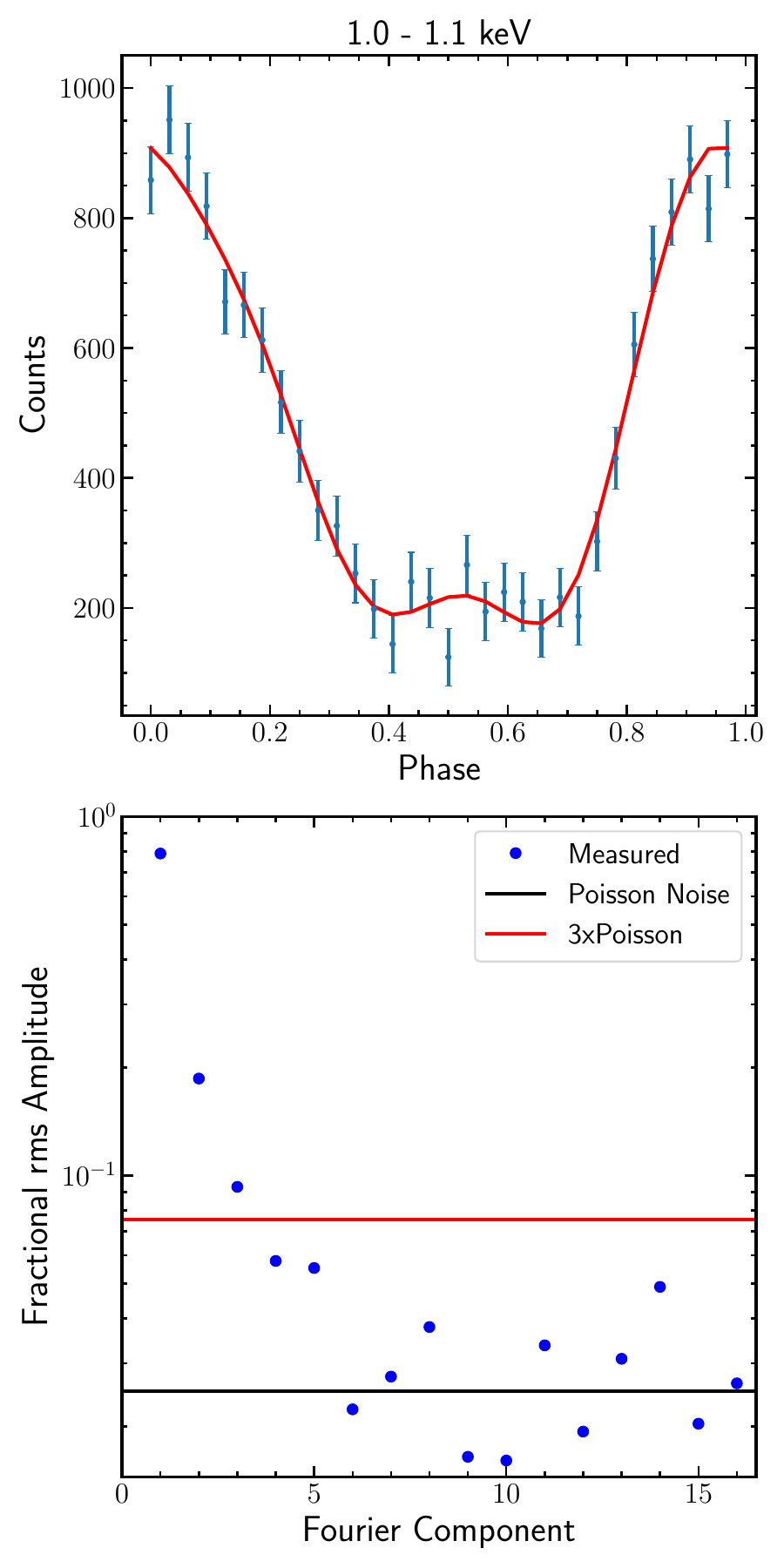}
    \caption{Same as \autoref{fig:pp1} but for \psrtt. Beyond 1.3~keV, the subtraction of the background results in no significant number of source counts.}
    \label{fig:psr1231pp}
\end{figure}

In \autoref{fig:psr1231pp}, we show examples of pulse profiles obtained with \nicer at several energy bands for \psrtt. For this pulsar, the background, estimated using \emph{scorp\_unc\_nsatmos2.txt} 
 file \citep{2024ApJ...976...58S}, dominates in multiple energy bands; beyond 1.3~keV, subtraction of the background results in no significant number of source counts. Up to three Fourier components are detectable in some of the energy bands. 

\begin{figure}
    \centering
    \includegraphics[scale=0.4]{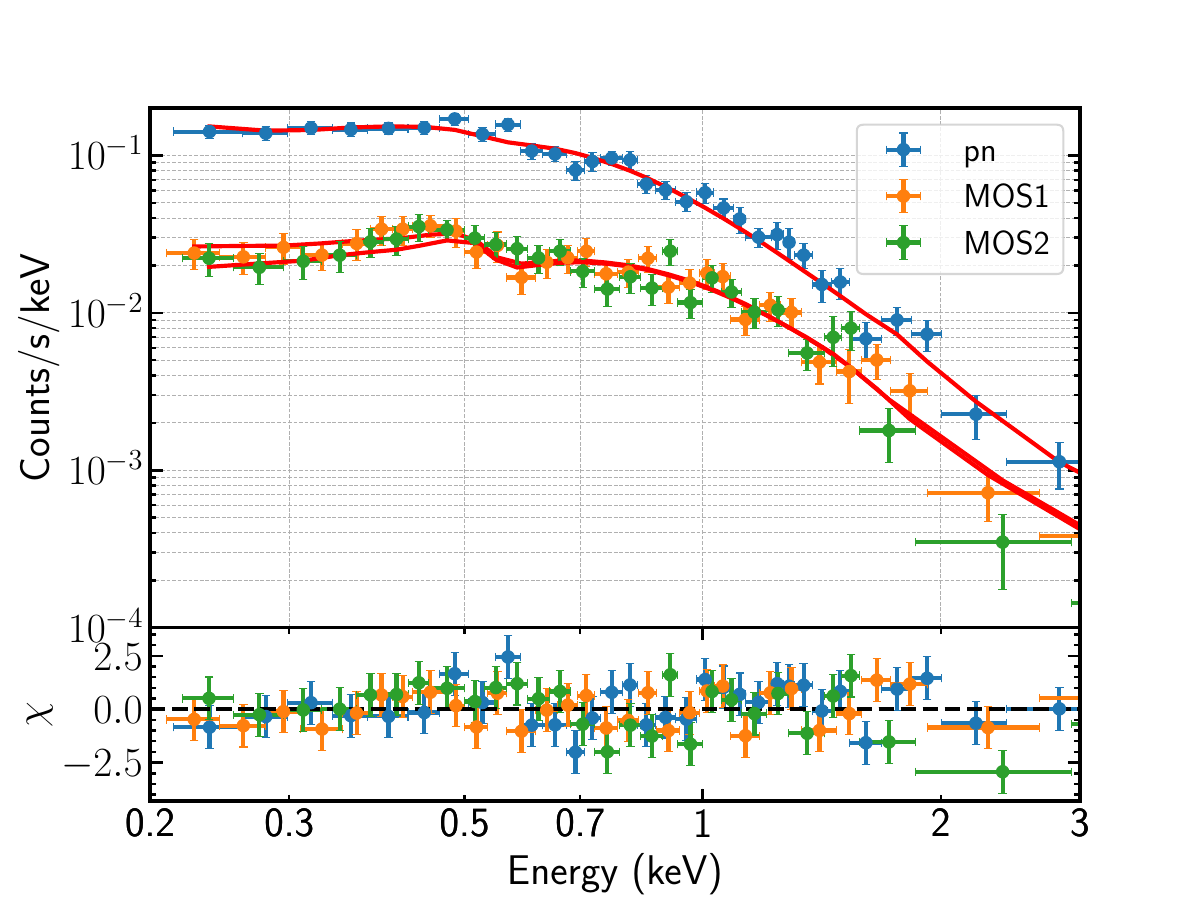}
        \includegraphics[scale=0.4]{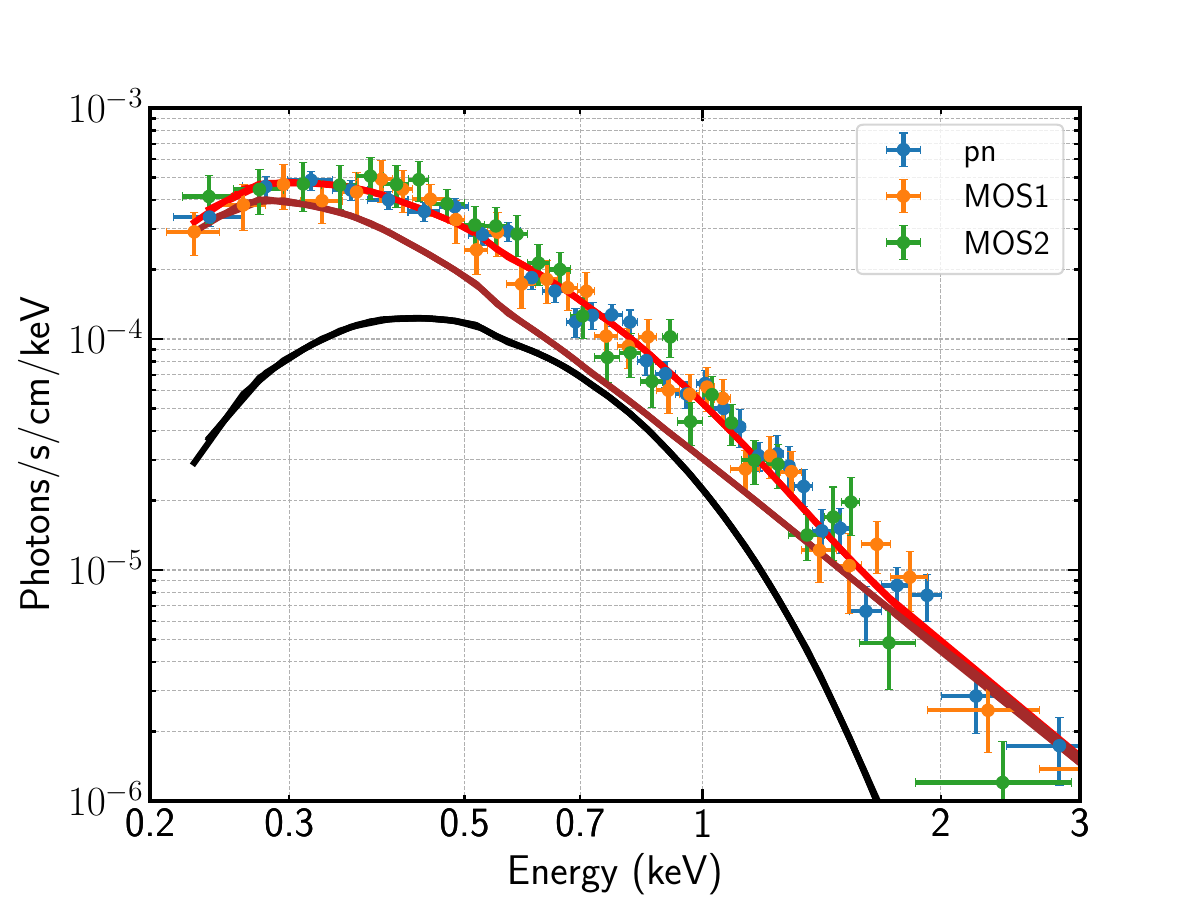}
    \caption{X-ray spectrum of \psrtt obtained with \xmm. The left panel shows the data and the best fit model (upper panel) together with the residuals (lower panel). The right panel shows the model components, which contain an absorbed blackbody (black) and a power-law component (red). }
    \label{fig:psr1231sp}
\end{figure}

\autoref{fig:psr1231sp} shows the X-ray spectrum obtained with \xmm, as well as the best-fit model. The spectral parameters are obtained from \cite{2017MNRAS.466.2560S} and \cite{2020MNRAS.498.2734B}, which also included {\it Chandra} and SUZAKU data. The one thermal component is significantly subdominant (up to a factor of 5) to the power-law component at all energies. Because of this, it is unclear what fraction of the pulsations are attributable to the surface thermal components. Fitting this dataset with two blackbodies results in large residuals towards high energies and an apparent emitting radius for one of the thermal components that is as small as 49~m, assuming a distance of 420~pc \citep{2017ApJ...835...29Y}.

\end{document}